\documentstyle[12pt,draft]{article}
\def\appendix{\par
 \setcounter{section}{0}
 \setcounter{subsection}{0}
 \setcounter{equation}{0}
 \def\theequation{Appx.~\arabic{equation}}
 \def\thesection{Appendix~\arabic{section}}}
\makeatletter
\def\vereq#1#2{\lower3pt\vbox{\baselineskip1.5pt \lineskip1.5pt
\ialign{$\m@th#1\hfill##\hfil$\crcr#2\crcr\gets\crcr}}}

\def\eqalign#1{\null\vcenter{\def\\{\cr}\openup\jot\m@th
  \ialign{\strut$\displaystyle{##}$\hfil&$\displaystyle{{}##}$\hfil
      \crcr#1\crcr}}\,}
\makeatother
\author{S.~S.~Sannikov-Proskurjakov\\
\it NSC Kharkov Institute of Physics and Technics,
Kharkov 310108,\\ \it 1 Academichna St., Ukraine\\
A.~A.~Stanislavsky\\
\it Institute of Radio Astronomy NASU, Kharkov 310002,\\ \it 4
Chervonopraporna St., Ukraine, E-mail:alexstan@ira.kharkov.ua\\
M.~J.~T.~F.~Cabbolet\\
\it Cavallilaan 119 5654 BC Eindhoven, The Netherlands,\\ E-mail:
m.cabbolet@wxs.nl}
\title{Elementary Particles in a New Quantum
Scheme}
\begin{document}
\large\tolerance8000\hbadness10000\emergencystretch3mm
\maketitle
\def\bbar #1{\bar {#1\vphantom{\bar #1}}}
\begin{abstract}
       Proceeding from the main principles of the non-unitary
quantum theory of relativistic bi-Hamiltonian systems, a system of
Lagrangian fields characterized by a certain dispersion law (mass
spectrum of particles), interactions between them and their
coupling constants are constructed.  In this article the mass
spectrum formula for ``bare'' fundamental hadrons is introduced,
and an a priori normalization of particle fields is found as well.
Numerical values of some parameters of the present theory are
determined.
\end {abstract}

\section {Introduction}
\thispagestyle {empty}
\begin{quote}
{\small\em Any physical  theory has to be expressed in an adequate
mathematical representation. Eventually the most of temporary
physical theories (even if they have been honored with the Nobel
prize) will be rejected because they are semiempirical. But the
present theory will survive untill a mathematically thinking
humanity exists, because it is an adequate mathematical theory.
Mathematics do not die, they are always applicable in a proper
place, and are only perfected} \end{quote}

      Apparently, to construct an adequate  theory of elementary
particles, it will be necessary to agree with a radical break-up
of our former representations about the nature of particles, that
are, to a large extent, based on geometrical concepts and
differential topology (in this relation the changeover in particle
theory should be certainly not less radical than formerly in
atomic theory).  First of all this remark is to be placed in the
context of the field concept, which, of course, loses force in the
subject area of physical processes on supersmall distances.

     Like it was necessary in the  construction of the consistent
atomic theory to completely reject the concept of a geometrical
trajectory of an electron and to assume another geometrical image
(field or de Broglie waves), now the very moment has come to
reject the field concept as functions on the space-time
continuum (see \cite{1}).

      It is important to emphasize that the transition from
trajectory to wave function is, as opposed to the transition from
point particle to vibrating string or membrane, not just a
generalization of the theory outgoing from a mechanistic analogy
with optics, but also it demonstrated a structure like a
differentiable manifold in space; in these terms others do not
exist \cite{1}. The transition from classical mechanics to quantum
mechanics is, in fact, the transition from Newtonian equations for
trajectories to the Schr$\rm\ddot o$dinger equations for wave
functions. Both theories are geometrical in character and are
based on the Newtonian concept of space as a differentiable
manifold:  in the first case the theory is formulated in terms of
the space variables (coordinates $x$); in the second case in terms
of the vector fiber bundle being built over the space (of bundle
sections or fields $\psi (x)$, see \cite{1}). In the development
of quantum mechanics, verifying  the sense of fields $\psi (x)$ as
waves of probability, the accuracy of the information that
followed directly from the mathematical apparatus of the theory
proved to be a refinement compared to the predictions of classical
theory. Based on this example a progress in fundamental physics is
supposed to be connected with the development and application of a
new and more perfect mathematical theory.\footnote {Let us add to
this that before the creation of quantum theory, attempts to
geometrize matter (relating the existence of matter to the
energy-momentum tensor\cite{2}) did not seem to be doomed. At
present, many people consider this approach to be outdated or even
erroneous. It follows that instead attempts should be made to
derive space from matter.}

      The Heisenberg-Schr$\rm\ddot o$dinger quantum theory has a
wide field of applications forming a set of  subareas, which are
in some sense enclosed in each other due to the characterizations
by different energy and/or distance levels (solids, molecules,
atoms, nuclei, elementary particles). To continue this ladder
further by assuming (without any solid basis) that particles
consist of quantum objects (quarks, gluons), that in turn consist
of preons, is a questionable way of thinking. The dubiousness of
that assumption is underlined by the fact that these objects are
unobservable in space (also, the phenomenon of quark confinement is
inconsistent with the concept of Lagrangian  or Hamiltonian
quantum systems). We think that quarks having small masses may not
exist inside protons (like electrons may not exist inside nuclei).

      In exactly this place radical changes seem to be required
in representations of the nature of particles: we should reject
composite  models of particles, being inadequate for the
description of the particle nature, and switch to another
paradigm: a new mechanistic concept, which is described by a
non-standard dynamical system coming to light on supersmall
distances,  where, as research shows \cite{1}, space-time can be
regarded as a discontinuum instead of a continuum. It is essential
that this statement is not an axiom, but a theorem:  then the
theoretical generalization has forced a well-founded possibility.
We consider  on supersmall distances the field concept to be
invalid, so that it makes no sense to speak about particles having
masses $\sim 10^{15}$ GeV.

      The transition to the new mechanics (dynamics of relativistic
bi-Hamiltonian system \cite{1}) goes together with a giving up of
differential equations in space-time. The new mechanics are
algebraic in character.

      If before  an action quantum $h$ was a fundamental value
needed for the transition from trajectories to fields, then now a
third constant $k$ with the dimension $cm^{-1}$ is the fundamental
value for a transition from continuum to discontinuum and from a
Lagrangian field system (elementary particles) to the relativistic
bi-Hamiltonian system.

      In the present article it will be shown that one of main
assignments of the new dynamical system is to create fundamental
particle fields.  In this relation the considered system (by
nature a two-level system) is analogous to an atom, in which
transitions generate photons. Particles, too, arise as a result of
quantum transitions in the above mentioned dynamical system with
an upper level (the upper half of light cone) and a lower level
(the lower half of light cone, see the main text of the article).
Thus, where the composite models tried to answer the question
($\underline{of\ what\ do\ fundamental\ particles\ consist}$), the
present theory answers another question:  $\underline {how\ to\
arise\ particles}$.  Having formulated the subject in such a way,
the topic is specific for dynamical theory (therefore for the sake
of brevity we shall call the proposed theory a dynamical one).

      The purpose  of this paper is to formulate the main laws
responsible for creating particles. In this perspective it is
reminded that Heisenberg already formulated the conditions an
adequate particle theory has to satisfy, in the form of main
questions that should be answered \cite{3}. Above all the theory
should explain:
1) the real observable mass spectrum of particles;
2) their symmetry properties;
3) the kinds and constants of interactions (particle charges).

      1) In the given work first  of all we determine explicit
states of the relativistic bi-Hamiltonian system --- semispinor
fields $f(x),\ \dot f(\dot x)$ called quanta $f$ and $\dot f$ in
another way.  At the same time the mathematical apparatus of the
new quantum theory, based on the extended Fock representation of
the Heisenberg algebra $h^{(*)}_8$ advanced in \cite{4}, plays an
important role. This part of the theory, following Neumann's
terminology \cite{5}, will be called the non-unitary quantum
theory I.

      Then we investigate amplitudes of the transition $f\to\dot f$
(this part of the theory will be  named the quantum theory II),
and derive equations which they satisfy. The relation of
fundamental particle fields $\psi^\Sigma (X,Y)$ with
amplitudes of the transition  $O^\Sigma (X,Y)$ is established, and
the mass spectrum formulae for ``bare'' fundamental hadrons and
leptons are defined.  In the present theory the particle fields
represent bilocal fields (the bilocality is a direct consequence
of the semispinor structure of particle fields or the
bi-Hamiltonian property of the considered dynamical system), which
describe non-point (smeared (non-infinitesimal)) objects: 1) four
coordinates $X_\mu$ of the affine space ${\bf A}_{3,1}$ describe
the motion of particles in our space-time; b) additional four
coordinates $Y_\mu$ describe a particle's interior as a object in
the second (internal) space-time ${\bf R}_{3,1}$ not depending on
our (external) space-time.  In the latter any movements are
possible only with speeds smaller than the velocity of light $c$.
Variables $Y_\mu$ are called hidden coordinates (as opposed to
open ones $X_\mu$).

      It is important to emphasize that the  quantum transition
$f\to\dot f$, irreversible by nature (since the structure
$\langle\dot f, f\rangle$ is non-Hermitian\cite{1}, it should be
noted that such structures are not suitable  for calculations of
probabilities), occurs when quanta $f$ (that might be identified
with Feynman's partons, providing an alternative to quarks!) are
in a hardly compressed state. Under these circumstances the quanta
$f$ form an ensemble, which is codetermined by the statistical
properties of the given ensemble (together with the ensemble of
quanta $\dot f$ that they pass into), that are described by the
distribution function in the relativistic Juttner form \cite{6}.
Here parameters (calculable in the theory) appear such as the
temperature of quanta $f\ \ T_f$, the temperature of quanta $\dot
f\ \ T_{\dot f}$ as well as their product --- the parameter
$\mu^2=T_f T_{\dot f}$ (the theory of ensembles will be called the
quantum theory III). The issue regarding the mass spectrum of
fundamental particles is closely related to the semispinor
structure of their fields, and in the hadron case it is to the
Gibbs distribution function of quanta $f$.

      Fields of fundamental hadrons (in particular, their a priori
normalization) and their  mass formulae (that exist of two
branches: baryonic and mesonic) are essentially defined by the
statistics of quanta $f$. To create leptons (in which case the
system passes from the upper half of light cone into the vertex of
cone, therefore $T_{\dot f}=0$ and $\mu^2=0$) the statistical
properties of quanta $f$ are not important. As a consequence the
lepton spectrum is much narrower than the hadron spectrum. In this
way, the existence of two families of particles --- hadrons and
leptons --- is connected with two topological different
transitions of the dynamical system: from the upper half of the
light cone either into the lower half (hadrons) or into the vertex
of the cone (leptons).

      Another limit case $\mu^2\to\infty$, having only sense for
fermions, results in objects characterized by a large imaginary
mass. Such objects have no wave function, therefore they do
not exist as particles in our spacetime (coordinate $X$). They
represent particles in the second spacetime (coordinates $Y$), and
show themselves only on supersmall distances as virtons playing
an essential role in the description of weak interactions (see
\cite{7}).

      2) For nearly all performed calculations about particles,
the relativistic bi-Hamiltonian system was used in a model based
on the algebra $h^{(*)}_8$.  In this model the isotopic symmetry
is represented by the group $U(1)$, all particles here are
electrically neutral, therefore the given model is not realistic.
In the realistic model based on the algebra $h^{(*)}_{16}$, the
isotopic symmetry is represented by the group $U(2)$. The Lorentz
symmetry of particles in both models is identical. It is connected
with the symmetry of the manifold of coordinates $X$ representing
the affine space ${\bf A}_{3,1}$.  Thus, in the present theory, a
space-time continuum in the form of ${\bf A}_{3,1}$ is formed
after the appearance of particle fields $\psi (X,Y)$ and switching
to interactions. If before first a space was mathematically
defined, whereon thereafter  functions were considered, then now
in the present theory this is reversed: we come to the idea of a
space after (to start from consideration of transition matrix
elements $\langle\dot f(\dot x),f(x)\rangle$) a certain set of
functions $\psi(X,Y)$ has been constructed where
$X=\frac{1}{2}(x+\dot x)$, $Y=\frac{1}{2}(x-\dot x)$ (for the
explanations of these formulae see the main text of the article).

      Results for the mass spectrum formulae and the a priori
normalization of the wave function of a particle are given for
both the realistic and the nonrealistic model.

      3) Particles arising from the quantum transition $f\to\dot f$ are
called fundamental. When considering a continuous degeneration of states
$\dot f$, which is described in the model $h^{(*)}_{16}$ by the
degeneration group $U_i(2)\otimes U_\ell (1)\otimes\dot T_{3,1}$, this
results in couplings between fields of fundamental particles ---
interactions are realized by fields corresponding to parameters of
the degeneration group (in the way of the Yang-Mills theory
\cite{7}). In accordance with three kinds of degeneration there
are three kinds of interactions and three kinds of quanta of
degeneration fields: strong ($U_i(2)$, first of all these are
$\eta$ and $\pi$-mesons), electromagnetic ($U_\ell (1)$, photons)
and gravitational ($\dot T_{3,1}$, gravitons) ones. In the present
theory the weak interactions are by nature distinguished from the
above mentioned ones, and, being given by correlations between
various fibers of the fiber bundle,  of a completely different
non-gauge kind \cite{7}.  Other kinds of particles (composite
ones) are connected with the latter interaction. The association
of $U_i(2)$-multiplets of fundamental particles with composite
particles results in higher symmetries (without quarks): ---
$SU(3), SU(4)$ (etc.)-multiplets.

      We would find the wrong (Planck) constant of electromagnetic
interactions, if we would determine the Heisenberg  algebra
$h^{(*)}_{16}$ by the commutation relations
\begin{equation}
[\phi_{\alpha k}\,,\bar\phi_{\beta m}]=
\delta_{\alpha\beta}\delta_{km}\,,
\quad [\phi_{\alpha k}\,,\phi_{\beta m}]=
[\bar\phi_{\alpha k}\,,\bar\phi_{\beta m}]=0
\label{eq11}
\end{equation}
More general relations are written in the form
\begin{equation}
[\phi_{\alpha k}\,,\bar\phi_{\beta m}]=
\Lambda\delta_{\alpha\beta}\delta_{km}
\label{eq12}
\end{equation}
(others are zero, as well as in (\ref{eq11})) where $\Lambda$ is
a dimensionless constant (analogous to the Planck constant) which
can impossibly be any number, and is connected with the
Lie algebra dimension of dynamical variables of our system,
equal to 136, by the formula $\Lambda=\sqrt{136}$.
 In the theory the ``bare'' electromagnetic charge of a particle is equal to
$e=\sqrt{\frac{hc}{136}}$. Radiative corrections to it are
considered in \cite{8}.

      In the brief exposition given below it is assumed that readers will
spare no efforts  to reconstruct the omitted calculations, filling
in blanks in the reasoning. For ourselves we have performed the
following task:  to state the main results of the theory as
coherently as possible in the given volume, and to show that there
is a class of indisputable mathematical trues deserving attention.
Probably the appendices will help to fill in some blanks.

\section[States of relativistic bi-Hamiltonian system \dots]
{States of relativistic bi-Hamiltonian system $f$ and $\dot f$
(Non-unitary quantum theory I)}\label{kd2}
      All symmetry and dynamical properties of particles
(including  the symmetry properties of space-time) are determined
completely by the defined properties of our dynamical system, in
particular, by the properties of its states $f(x)$ and $\dot
f(\dot x)$ satisfying equations (see \cite{1}):
\begin{equation}
-i\frac{\partial}{\partial x_\mu}f(x)=p_\mu f(x),\quad
-i\frac{\partial}{\partial\dot x_\mu}\dot f(\dot x)=
\dot p_\mu\dot f(\dot x).
\label{eq21}
\end{equation}
Here $x_\mu$ and $\dot x_\mu$ are coordinates of the translation group
$T_{3,1}$ and $\dot T_{3,1} $.  The generators of that group are the
operators $p_\mu$ and $\dot p_\mu$, written in the extended Fock
representation in the form (considering the model $h^{(*)}_8$, see
\cite{4}):
\begin{equation}
P_\mu=kh\ \bbar\varphi\stackrel {+}{\sigma}_\mu\varphi\,,\quad \dot
p_\mu=-kh\ \frac{\partial}{\partial\varphi}\stackrel {-}
{\sigma}_\mu\frac{\partial}{\partial\bbar\varphi}
\label{eq22}
\end{equation}
(now the variables $z_\alpha\,,\,\bbar z_\alpha\,,\,\alpha=1,2$, used in
\cite{4}, are denoted by $\varphi_\alpha\,,\,\bbar\varphi_\alpha$, and
additional variables by $\varphi\dot =\varphi_2\,,\,\bbar\varphi\dot =
\bbar\varphi_2$). $f(x)$ and $\dot f(\dot x)$ represent fields
not defined on space-time (hence, these are not Lagrangian
fields), but on the groups $T_{3,1}$ and $\dot T_{3,1}$
respectively, which act in a fiber ($x_\mu\,,\,\dot x_\mu$ are
coordinates in fiber, see \cite{1}). Since 4-vectors $p_\mu$ and
$\dot p_\mu$ are isotropic:  $p^2_\mu= \dot p^2_\mu=0$, the fields
$f(x)\,,\,\dot f(\dot x)$ describe objects with zero mass. We will
call them simply quanta $f$ and $\dot f$. The solutions of
equations (\ref{eq21}) are written in the form
$f(x)=e^{ipx}f_0\,,\,\dot f(\dot x)=e^{i\dot p\dot x}\dot f_0$.

      1) The states $\dot f_0$ are defined as solutions of stationary
equations
\begin{equation}
\dot p_\mu\dot f_0=\rho_\mu^{(0)}\dot f_0
\label{eq23}
\end{equation}
and belong (as states of the physical vacuum \cite{4}) to the space of
additional variables  $\dot{\cal F}_0$. Such solutions are written
in the form
\begin{equation}
\dot f_0=\dot f_z=Ce^{\bbar
z\varphi-\bbar\varphi z}= Ce^{2i\,\Im\!\mbox{\scriptsize\it
m}\,\bbar z\varphi} \label{eq24}
\end{equation}
where $C$ is a  normalization constant and $z$ is a numerical
(complex) parameter by means of which eigenvalues $\rho_\mu^0$ in
(\ref{eq23}) are written in the form $\rho_\mu^0=\bbar
z\!\!\stackrel{-}{\sigma}_\mu\!\!z=\vert z\vert^2(0,0,-1,-i)$
where $z={0\choose z}$ (therefore we chose to use the unit system
$c=h=k=1$). It follows from this that ${\rho_\mu^0}^2=0$, and
$\rho_0^0\leq 0$ (the latter condition is a consequence of the
boundary condition put on $\dot f_z$, because of the
limitation $\dot f_z$ for $\vert\varphi\vert\to\infty$), so that a
4-momentum of quanta $f$ belongs to the lower half of the light
cone. Since $\varphi\,,\,\bbar\varphi$ are Lorentzian scalars,
$\dot f_z$ is a  relativistic invariant value. Such states show
hardly any degeneration of energy-momentum $\rho_\mu$. In fact,
performing the transformation $T(v)$ on equation (\ref{eq23}),
where $v\in SL(2,{\bf C})$, we come to the equations $\dot
p_\mu\dot f_0=\rho_\mu\dot f_0$ in which
$\rho_\mu=(L^{-1}(v))_{\mu\nu}\,\rho^{(0)}_\nu$ can accept
arbitrary values  from the lower half of light cone: $\rho\in
N_-$. In this case the  vertex of the cone $\rho=0$ (with the
parameter $z=0$) forms an invariant subset in respect to $L(v)$,
so that the set of all values of the 4-momentum $\rho_\mu$
decomposes in two invariant subsets: open $N_-$ and closed $\{0\}$
ones.

      The solution (\ref{eq24}) can be written in the form
\begin{equation}
\dot f_z=Ce^{i\sqrt{-{\cal P}^2}\sin\nu}
\label{eq25}
\end{equation}
where $-{\cal P}^2=2\pi_\mu\rho_\mu$, and $\nu=\arg\varphi-\arg
z$.  Really, since by definition
$\pi_\mu=\bbar\varphi\stackrel{+}{\sigma}_\mu \varphi$,
$\pi_\mu\rho_\mu=\bbar\varphi\stackrel{+}{\sigma}_\mu\varphi\
\bbar z\stackrel{-}{\sigma}_\mu z=2\vert\bbar z\varphi\vert^2$
where $\varphi\dot =\varphi_2$~.

      As a Lorentz-invariant  measure for $N_-$ (someway
normalized), we have
\begin{equation}
d\mu_{\dot f}=\frac{2}{\pi}\,\theta (-\rho_0)\,\delta
(\rho^2)\,d^4\rho\,,
\label{eq26}
\end{equation}
and on $\{0\}$ the measure is
\begin{equation}
d\mu_0=\delta^4(\rho)\,d^4\rho\,.
\label{eq27}
\end{equation}
$d\mu_{\dot f}$ and $d\mu_0$ are called the measures of final states
$\dot f$.

      2) Next, since $\pi_\mu=\bbar\varphi\!\!\stackrel{+}
{\sigma}_\mu\!\!\varphi$, $\pi_\mu^2=0$, and
$\pi_0=\bbar\varphi_\alpha \varphi_\alpha>0$ so  that $\pi\in N_+$
is the upper half of the light cone.  By means of  variables
$\pi_\mu$, a Lorentz-invariant measure for the Lagrangian plane
$\prod_{\alpha=1,2}\frac{i}{4\pi}\,d\varphi_\alpha\,\wedge\,
d\bbar\varphi_\alpha$\cite{4} can be written in the form $\theta
(\pi_0)\,\delta (\pi^2)\,d^4\pi\,d\nu$. The measure
\begin{equation}
d\mu_f=\frac {1}{(2\pi)^{3/2}}\,\theta (\pi_0)\,\delta (\pi^2)
\,d^4\pi\,\frac{d\nu}{2\pi}
\label{eq28}
\end{equation}
concentrated on the upper half of the light cone $N_+$,
normalized in a reasonable way, will be called the measure of
initial states $f$. We now notice that states $f_0$ are not in the
least defined by equations (\ref{eq21}).

      In defining $f_0$, the operator $\stackrel{\wedge}{M^2}=
2\dot p_\mu p_\mu$ (operator of square mass) and its stationary group
$G_{\stackrel {\wedge}{M^2}}=GL_\ell (2,{\bf C})\otimes U_i(1)\otimes
H_i(1)$ play an important role. Here $GL_\ell (2,{\bf C})=SL_\ell
(2,{\bf C})\otimes U_\ell (1)\otimes H_\ell (1)$. In respect to the
separation of phase transformations $U(1)$ and $H(1)$ on $U_\ell (1),\
H_\ell (1)$ and $U_i (1),\ H_i(1)$ see \cite{3}.

      The states $f_0$ are defined as eigenfunctions of the operator
$\stackrel{\wedge}{M^2}$:
\begin{equation}
\stackrel{\wedge}{M^2}\!\! f_0^\Sigma=F^0_\Sigma f_0^\Sigma\,,
\label{eq29}
\end{equation}
forming a finite-dimensional multiplet $\Sigma$ of the group
$G_{\stackrel{\wedge}{M^2}}$ (in this case the group $G_{\stackrel
{\wedge}{M^2}}$ plays the role of degeneration group for the
``Hamiltonian'' $\stackrel {\wedge}{M^2}$). Here $\Sigma$ is a
set of quantum numbers $\Sigma=(s,F,D;Y,N)$ which is defined
as a finite-dimensional irreducible  representation of the group
$G_{\stackrel{\wedge}{M^2}}$ ($s$ is the Dirac spin connected
with the group $SL_\ell (2,{\bf C})$, $F$ and $D$ are the fermion
charge and dilatation connected with $U_\ell (1)$ and $H_\ell
(1)$, and $Y$ and $N$ are the hypercharge and isotonic quantum
number connected with $U_i(1)$ and $H_i(1)$). Such states,
obviously, are written in the form $f_0^\Sigma=O^\Sigma
(\varphi_\alpha\,,\bbar\varphi_\beta\,; \varphi\,,\bbar\varphi)$
of an homogeneous polynomial of degree $N$ in the variables
$\varphi_\alpha\,,\bbar\varphi_\beta\,,\varphi\,, \bbar\varphi$
and one of degree $D$ in the variables $\varphi_\alpha\,,
\bbar\varphi_\beta$. In this case the number $Y$ is the difference
between the number of variables $\varphi_\alpha\,,\varphi$ and
$\bbar\varphi_\beta \,,\bbar\varphi$. The number $F$ is the
difference between the number of variables $\varphi_\alpha$ and
$\bbar\varphi_\beta$, that have equal values for the whole multiplet.
The number $S=Y-F$ is called strangeness (in this perspective
the additional variables $\varphi,\bbar\varphi$ can possibly also
be named quanta of strangeness; the name is justified by an
adherence of a variable $\bbar\varphi$ to a normal spinor
$\varphi_\alpha$, and subsequantly comparing the strange spinor
$\varphi_\alpha\bbar\varphi$, with spurions in
Heisenberg's theory \cite{3}). For states $O^\Sigma (\varphi)$
eigenvalues $F^0_\Sigma$ of the operator $\stackrel {\wedge}{M^2}$
are equal (see \ref{dix1}) to
\begin{equation}
F^0_\Sigma=-\left (N^2-Y^2+8N+16\right)=-\left[(N+4)^2-Y^2\right].
\label{eq210}
\end{equation}
Polynomials $O^\Sigma(\varphi)$ will be called skeletons of particles.

      From the solutions of $f_0^\Sigma (x)=e^{i\pi x}\,O^\Sigma
(\varphi)$ it is possible to construct coherent states --- fields
\begin{equation}
O^\Sigma (x)=\int f_0^\Sigma (x)\,d\mu_f=\frac{1}{(2\pi)^{3/2}}\,
\int e^{i\pi x}\,\theta (\pi_0)\,\delta (\pi^2)\,d^4\pi\,O^\Sigma
(\pi)
\label{eq211}
\end{equation}
where
\begin{equation}
O^\Sigma (\pi)=\frac{1}{2\pi}\,\int d\nu\,O^\Sigma(\varphi)\,.
\label{eq212}
\end{equation}
These (massless) fields exist on the group $T_{3,1}$, and should
not be mixed up with Lagrangian fields given on
space-time.

      $f^\Sigma_0(x)$ or $O^\Sigma (x)$ are states of an isolated
quantum  $f^\Sigma$. In case a quantum $f$ forms an ensemble with
quanta $f$ (in other fibers) it is necessary to take into account
statistical properties of the ensemble. Usual statistical
reasonings (see, for example, \cite{9}) result in consideration of
the distribution function of quanta $f$ with energy
$\pi_0=\bbar\varphi_\alpha\varphi_\alpha$ (in the frame of reference
connected with the space ${\cal F}_0$, it will be $\bbar\varphi
\varphi$), that, representing the Gibbs distribution function, is
written in the form
\begin{equation}
w_f=\exp\left(-\frac{\bbar\varphi\varphi}{T_f}\right)
\label{eq213}
\end{equation}
where $T_f$ is the temperature of the ensemble of quanta $f$.
By reasons outlined in \cite{1}, it follows that quanta $f$ (which
henceforth will be identified with Feynman's partons) arise from a
(generally speaking, reversible) phase transition ``particles
$\Leftrightarrow$quanta $f$'' , that occurs at superhigh densities
of particles (for example, at the Universe collapse or in collisions
of particles with superhigh energies), being in thermal balance
with particles (at the temperature $T_f$). In this sense they are
similar to photons (being  in thermal balance with the walls of a
black body) and are characterized by minimum of free energy $F_f$,
i.\ e.\ $\left(\partial F_f/\partial N_f\right)_{T_f\,,\,V}=0$
where $N_f$ is the number of quanta $f$ in the ensemble which
takes a volume $V$.  This condition, as is known \cite{9}, entails
the vanishing chemical potential of quanta $f$ so that in the case
of quanta $f$ the free energy $F$ coincides with the
thermodynamical potential $\Omega$.

      Clearly, the ensemble of quanta $f$ can be also considered as
a gas. Since in each fiber there is not more than one
copy of our dynamical system, and since the number of fibers is
denumerable, in contrast to the nondenumerably many points that
form the desintegrated space (so that, at the phase transition
``continuum $\to$ discontinuum '' the majority of points are
empty, hence, occupation averages are $\bbar n_f\ll 1$), the Gibbs
distribution coincides in effect with the Boltzmann distribution
(and in the strongly degenerate limit, which is never reached for
reasons indicated below, it coincides with the Fermi or Bose
distributions depending on statistics of the skeleton $O^\Sigma
(\varphi)$).

      With  regard to the  statistical properties a field of
quantum $f$ in the ensemble is written in the form $f^\Sigma
(x)=w_f f^\Sigma_0$.

      At the moment of the  quantum transition $f\to\dot f$, which
will be considered below, the function $w_f$ will be written in
the relativistic invariant form, i.\ e.\ in the Juttner
distribution function form (see \cite{6})
\begin{equation}
w_f=\exp\left(-\frac{\pi\rho}{2\mu^2}\right)=
\exp\left(-\frac{-{\cal P}^2}{4\mu^2}\right)
\label{eq214}
\end{equation}
where by definition (which we already mentioned) $-{\cal
P}^2=2\pi\rho$, and the parameter $\mu^2=3T_fT_{\dot f}$ where
$T_{\dot f}=\frac{1}{3}\vert z\vert^2$ denoted the temperature
of quanta $\dot f$. Really, since $\pi\rho$ is relativistic
invariant,  and since in the special frame of reference connected
with the space ${\cal F}_0$ in which $\rho_\mu=\rho^0_\mu$ and
$\varphi_\alpha={0\choose\varphi}$, we have $\pi\rho=2\vert
z\vert^2\bbar\varphi\varphi$. For $\bbar\varphi\varphi$ we obtain
$\bbar\varphi\varphi=\frac{\pi\rho} {2\vert z\vert^2}$. Thus, in
(\ref{eq214}) a 4-momentum of quanta $\dot f\ \rho_\mu$ takes
the part of 4-velocity. It should also be noticed that $w_f$ is a
homogeneous function on the group $T_{3,1}$, and therefore it
depends neither on any angular moment $\pi_\mu x_\nu-\pi_\nu
x_\mu$ nor on $x_\mu$.

      As a  result it  is possible to introduce the fields of
quanta $f$ in the ensemble so
\begin{equation}
f^\Sigma=w_f\,f_0^\Sigma=e^{-\frac{-{\cal P}^2}{4\mu^2}}\,
O^\Sigma (\varphi).
\label{eq215}
\end{equation}

      Similarly, statistical properties of the quantum ensemble
$\dot f$, being described by the distribution function
\begin{equation}
w_{\dot f}=\exp\left(-\frac{\rho_0 +\mu_{\dot f}}
{T_ {\dot f}}\right)\simeq\exp\left(-\frac{\mu_{\dot f}}
{T_ {\dot f}}\right)=\frac{1}{Z}
\label{eq216}
\end{equation}
in which  $\mu_{\dot f}$ is a chemical potential of quanta $\dot f$
(non-zero, since the quanta $\dot f$ are accumulated as a result
of the transition $f\to\dot f$, see below), are taken into
account, and $T_{\dot f}$ are their temperatures as defined above.
In (\ref{eq216}) we consider $\rho_0\ll\mu_{\dot f}$. As a result a
field of quanta $\dot f$ in the ensemble is written in the
form
\begin{equation}
\dot f_z=\frac{C}{Z}\,e^{i\sqrt{-{\cal P}^2}\sin\nu}.
\label{eq217}
\end{equation}

\section [Amplitudes of the transition \dots] {Amplitudes of the transition
$f\to\dot f$ (Non-unitary quantum theory II)}
      The amplitude of transition is understood to be a transition matrix
element averaging
\begin{equation}
\langle\dot f(\dot x),f^\Sigma (x)\rangle=\int\overline {\dot
f(\dot x)}\,f^\Sigma (x)\,d\mu_f
\label{eq31}
\end{equation}
with measure of final states $d\mu_{\dot f}$ or $d\mu_0$,
i.\ e.\ the value
\begin{equation}
O^\Sigma (X;Y)=\int d\mu_{\dot f,0}\,\langle\dot f(X-Y),
f^\Sigma (X+Y)\rangle=\langle\!\langle\,\dot f(X-Y),f^\Sigma
(X+Y)\,\rangle\!\rangle,
\label{eq32}
\end{equation}
where we denoted $X=\frac{1}{2}(x+\dot x),\ Y=\frac{1}{2}(x-\dot
x)$. Thus, there are two similar but topological different transitions:
the  transition from the upper half of the light cone (the measure
$d\mu_f$) to the lower (the measure $d\mu_{\dot f}$), that will be
named hadronic, and the transition from the upper half to the
vertex of the cone (the measure $d\mu_0$), that will be named
leptonic.  By the first transition heavy particles are created, by
the second low-mass particles.

      It is plain to see that the amplitudes of lepton transitions
differ  only from the coherent fields (\ref{eq211}) by the factor
$1/Z$, and are only non-zero if skeletons $O^\Sigma
(\varphi)=P(\pi)$ or $O^\Sigma
(\varphi)=\varphi_\alpha\bbar\varphi P (\pi)$ where $P(\pi)$ are
polinomials about $\pi_\mu=\bbar\varphi\stackrel
{+}{\sigma}_\mu\varphi$. Amplitudes appropriated for such skeletons
are written in the form ($x=X+Y$)
\begin{equation}
\psi^\Sigma (x)=P\left(-i\frac{\partial}{\partial x}\right)X^0(x),
\quad\psi^\Sigma (x)=P\left(-i\frac{\partial}{\partial x}
\right)\nu (x),
\label{eq33}
\end{equation}
where
\begin{equation}
\eqalign{
X^0(x)=\frac{1}{(2\pi)^{3/2}}\,\frac{1}{Z}\,\int e^{i\pi x}
\theta (\pi_0)\,\delta (\pi^2)\,d^4\pi\,,\\
\nu_\alpha (x)=\frac{1}{(2\pi)^{3/2}}\,\frac{1}{Z}\,\int
e^{i\pi x}\theta (\pi_0)\,\delta (\pi^2) \,\nu_\alpha (\pi)\,
d^4\pi\,,}
\label{eq34}
\end{equation}
and $\nu_\alpha (\pi)=\frac{1}{2}{\pi_1-i\pi_2 \choose\pi_0-\pi_3}=
\bbar\pi a$. Here $\bbar\pi=\ \stackrel {-}{\sigma}_\mu\pi_\mu=\pi_0+
\vec\sigma\vec\pi$, and $a=\frac{1}{2}{0\choose 1}$ is a constant
spinor. (\ref{eq33}) are special bilocal fields
representing gradient waves from fields $X^0 (x),\nu_\alpha (x)$.
From this it follows that only two kinds of particles arise from
leptonic transitions,
in the case of the model $h^{(*)}_8$ this is a massless boson
$X^0$ with spin 0 and a massless fermion $\nu (x)$ with spin 1/2.
The fields of particles satisfy equations: $\Box X^0=0,\,\stackrel
{+}{\sigma}_\mu\frac{\partial}{\partial x_\mu}\ \nu=0$. The bilocal
fields (\ref{eq34}) result from a translation on the 4-vector
$Y_\mu$ of a local field $\psi^\Sigma (X)$ (and herein they are
specific): $\psi^\Sigma (X+Y)=e^{Y\frac{\partial}{\partial X}}
\psi^\Sigma (X)$.

      Now we discuss hadron transitions. If to denote
${\cal P}=\pi-\rho\,,\,{\cal Q}=\pi +\rho$, amplitudes of hadron
transitions can be represented in the form
\begin{eqnarray}
O^\Sigma (X;Y)=\frac{C}{(2\pi)^{3/2}}\int e^{i{\cal P}X+i{\cal Q}Y}\,
\theta ({\cal P}_0+{\cal Q}_0)\,\theta ({\cal P}_0-{\cal Q}_0)\,\times
\nonumber\\
\times\ \delta ({\cal Q}^2+{\cal P}^2)\,\delta ({\cal P}{\cal Q})\,
O^\Sigma ({\cal P};{\cal Q})\,d^4{\cal P}\,\frac{d^4{\cal Q}}{2\pi}\,,
\label{eq35}
\end{eqnarray}
where
\begin{equation}
O^\Sigma ({\cal P};{\cal Q})=\frac {1}{Z}\,\int e^{-i\sqrt{-{\cal P}^2}
\sin\nu-\frac{-{\cal P}^2}{4\mu^2}}\,O^\Sigma (\varphi)\,\frac{d\nu}
{2\pi}
\label{eq36}
\end{equation}
(these values are calculated in Appendix~2). Let us act
on the amplitude $O^\Sigma (X;Y)$ (\ref{eq32}) by the D'Alember
operator with respect to variables $X_\mu$:
$\Box_X=-\frac{\partial^2}{\partial t^2}+
\frac{\partial^2}{\partial\vec X^2}$ and take into account
$p_\mu^2=\dot p_\mu^2=0$. We finally obtain
\begin{equation}
\Box_X O^\Sigma (X;Y)=\langle\!\langle\,\dot f(X-Y),\
\stackrel {\wedge}{M^2}\!\! f^\Sigma (X+Y)\,\rangle\!\rangle
\label{eq37}
\end{equation}
where $\stackrel {\wedge}{M^2}=2\dot p_\mu p_\mu$ is
the operator introduced into consideration above. Denote
$[\stackrel {\wedge}{M^2}~,e^{ipx}]e^{-ipx}=\stackrel
{\wedge}{\Im}(x)$. By definition we have $\stackrel
{\wedge}{\Im}(0)=0$. This condition allows specifically to isolate
a homogeneous part in the transition amplitude equation. We denote
\begin{equation}
\stackrel {\!\!\!\wedge}{\Im^\Sigma}(X;Y)=\langle\!\langle\,\dot
f(X-Y),\stackrel{\wedge}{\Im}(X+Y)f^\Sigma(X+Y)
\,\rangle\!\rangle.
\label{eq38}
\end{equation}
It is not difficult to make sure that
\begin{equation}
\stackrel{\wedge}{M^2}\!\!f^\Sigma=f_\Sigma(-{\cal P}^2)
f^\Sigma
\label{eq39}
\end{equation}
where (see \ref{dix1})
\begin{equation}
f_\Sigma (-{\cal P}^2)=F^0_\Sigma +(N+5)\frac{-{\cal P}
^2}{\mu^2}-\frac{\left(- {\cal P}^2\right)^2}{4\mu^4}\,,
\label{eq310}
\end{equation}
and $F^0_\Sigma$ is defined by formula (\ref{eq210}). In
accepted notations the equation (\ref{eq37}) is written in the form
of the inhomogeneous equation
\begin{equation}
\left(\Box_X-F_\Sigma (\Box_X)\right)O^\Sigma (X;Y)=\Im^\Sigma (X;Y).
\label{eq311}
\end{equation}
The latter can be written in the form of the Bopp equation (the
left-hand part in (\ref{eq311}) is a polynomial of second degree
regarding $\Box_X$)
\begin{eqnarray}
{\cal D}(\Box_X)O^\Sigma (X;Y)=\left(\Box_X-M^2_{\Sigma\,B}
\right) \,\times\nonumber\\
\times\ \left(\Box_X-M^2_{\Sigma\, M}\right)O^\Sigma (X;Y)=
4\mu^4\Im^\Sigma (X;Y)
\label{eq312}
\end{eqnarray}
in which $\Im^\Sigma(X;Y)$ (\ref{eq38}) takes the part of a source
for transition amplitudes $O^\Sigma(X;Y)$. In (\ref{eq312})
\begin{equation}
M^2_{\Sigma\,B}=2\mu^2\left\{N+5-\mu^2+\sqrt{\mu^4-2\mu^2(N+5)+
Y^2+2N+9}\right\},
\label{eq313}
\end{equation}
and
\begin{equation}
M^2_{\Sigma\,M}=2\mu^2\left\{N+5-\mu^2-\sqrt{\mu^4-2\mu^2(N+5)+
Y^2+2N+9}\right\}.
\label{eq314}
\end{equation}

\section{Fields of particles}
      As usual, a solution of the inhomogeneous equation (\ref{eq312})
can be written in the form of a sum $O^\Sigma=\psi^\Sigma +G^\Sigma$
of  the partial solution $G^\Sigma$ of the inhomogeneous equation
(\ref{eq312}) and  the general solution $\psi^\Sigma$ of the
homogeneous equation
\begin{equation}
\left (\Box_X-M^2_{\Sigma\,B}\right)\left(\Box_X-M^2_{\Sigma\, M}
\right)\psi^\Sigma(X; Y)=0.
\label{eq41}
\end{equation}
Solutions of equation (\ref{eq41}) have physical sense,
since they allow
\begin{equation}
\psi^\Sigma(X;Y) =\frac{1}{(2\pi)^{3/2}}\,\int e^{i{\cal P}X}\,
\psi^\Sigma ({\cal P};Y)\,\delta ({\cal P}^2+M^2_{\Sigma})\,d^4
{\cal P}.
\label{eq42}
\end{equation}
In the asymptotics $\vert X\vert\gg\vert Y\vert$ the solutions
(\ref{eq42}) pass to local fields observable in ${\bf A}_{3,1}$
\begin{equation}
\psi^\Sigma (X;0)=\psi^\Sigma(X)=\frac{1}{(2\pi)^{3/2}}\,
\int e^{i{\cal P}X}\,\theta ({\cal P}_0)\,\psi^\Sigma ({\cal P})
\,\delta ({\cal P}^2+M^2_{\Sigma})\,d^4{\cal P}
\label{eq43}
\end{equation}
where $\theta ({\cal P}_0)\psi^\Sigma ({\cal P})=\psi^\Sigma
({\cal P};0)$, which are characterized by the
certain dispersion law ${\cal P}^2=-M^2_\Sigma$. In quantum field
theory such fields are interpreted as point particles. As for
$\psi^\Sigma (X;Y)$, the more general fields are bilocal. Such
fields are interpreted as non-point (smeared) objects, their
interior is described  by coordinates $Y_\mu$ (comparing $X_\mu$
it can be said that these are usual coordinates of space-time). It
is important to emphasize that a particle's interior is not at all
part of our space-time ${\bf A}_{3,1} $, it owes a special space.
Note also that the signature and connectivity on manifold ${\bf
A}_{3,1}(\ni X_\mu)$ is completely defined by dynamical symmetry
of our system, i.\ e.\ by the symmetry of the algebra $h^{(*)}_8$.

      It is important to emphasize that equations (\ref{eq41})
are no evolutionary equations for fields $\psi^\Sigma$ (and
neither are the Klein-Gordon equations), they are only dispersion
equations (see below).

      As distinguished from $\psi^\Sigma$ the solutions $G^\Sigma$
have not any immediate physical interpretation as they are not
known in the standard quantum field theory.  In this perspective
we require the non-physical solutions $G^\Sigma$ to be orthogonal
to the physical fields $\psi^\Sigma$ in the sense of the
Stueckelberg scalar product
\begin{equation}
\int\psi^\Sigma
(X;Y)\,\overline{G^\Sigma(X;Y')}\,d^4X=0.  \label{eq44}
\end{equation} This condition permits to express the function
$\psi^\Sigma({\cal P};Y)$ in terms of the general solution
(\ref{eq42}) (in the case of hadrons). As
$G^\Sigma=O^\Sigma-\psi^\Sigma$, we have
\begin{displaymath} \int\psi^\Sigma (X;Y)\,\overline{\psi^\Sigma
(X;Y')}\,d^4X= \int\psi^\Sigma (X; Y)\,\overline{O^\Sigma (X;
Y')}\,d^4X.  \end{displaymath} Applying the Fourier transformation
on $\psi^\Sigma$ (\ref{eq42}) and on $O^\Sigma$ (\ref{eq35}) and
integrating over $d^4X$, we come to a relation that implies that
the equality \begin{eqnarray} \psi^\Sigma ({\cal P};Y)=\int
e^{i{\cal Q}Y}\,\theta ({\cal P}_0+ {\cal Q}_0)\,\theta ({\cal
P}_0-{\cal Q}_0)\times\nonumber\\ \times\ \delta ({\cal P}^2+{\cal
Q}^2)\,\delta ({\cal PQ})\, O^\Sigma ({\cal P};{\cal
Q})\,\frac{d^4{\cal Q}}{2\pi} \label{eq45} \end{eqnarray} must be
valid, and for that purpose the normalized constant $C$ in
(\ref{eq217}) or in (\ref{eq35}) must be equal to
\begin{equation}
C=\delta (0).
\label{eq46}
\end{equation}
As can be seen, the value of $C$ is infinite, and consequently all
values containing it as $\dot f_z(\dot x)\,,\,O^\Sigma
(X;Y)\,,\,\Im^\Sigma (X;Y)\,,\,G^\Sigma (X;Y)$ are infinite, and
hence there is no direct physical sense. This result is very
important for a consistent physical interpretation of the theory,
particularly because it permits to answer the question whether
final states (in case of hadron transitions) of our system $\dot
f$ are observable. Regarding these non-physical states the answer
is that they are not observable (after the transition $f\to\dot f$
and creation of particle fields $\psi^\Sigma$ the quanta $\dot f$
remain in the orthogonal adjunct to the field $\psi^\Sigma$).
Particle fields $\psi^\Sigma(X;Y)$ and fields $f^\Sigma(x)$ of
quanta $f$ (partons) are only finite (that consequently have
physical sense).

      As it is seen, the present theory is broader than quantum
field theory of particles, containing non-observable
(in terms of measure of our space-time) values on principles such as
$\dot f_z(\dot x)\,,\,O^\Sigma(X;Y)\,,\,\Im^\Sigma (X;Y)\,,\,G^\Sigma
(X;Y)$ going beyond the class of generalized functions (see \cite{1}).

      Next, focussing the attention on the structure of bilocal
fields $\psi^\Sigma (X;Y)$, it is possible to see that it admits
a representation in the form of a smeared local field $\psi^\Sigma
(X)$, i.\ e.
\begin{equation}
\psi^\Sigma(X;Y)=F_\Sigma\left(Y;-i\frac{\partial}{\partial X}
\right)\,\psi^\Sigma (X)
\label{eq47}
\end{equation}
where $F_\Sigma\left(Y;-i\frac{\partial}{\partial X}\right)$ is a
smearing (differential) operator. This operator depends on the
kind of a particle $\Sigma$, in particular, on spin
of a particle $s$ as well as on a spin projection (as
$O^\Sigma({\cal P};{\cal Q})$ depends on $Q$). However in
certain calculations it satisfies to use an approximate
expression for $F_\Sigma$, approximately equal for particles of
any kind $\Sigma$, namely
\begin{equation}
F(Y;{\cal P})=\frac{1}{2\pi}\,\int e^{i{\cal Q}Y}\,\delta
({\cal Q}^2+{\cal P}^2)\,\delta ({\cal PQ})\,d^4{\cal Q}
\label{eq48}
\end{equation}
(therefore being valid).

      Since for a freefalling particle ${\cal
P}_\mu$ is a time-like 4-vector, ${\cal Q}_\mu$, as it follows
from (\ref{eq48}), is a space-like 4-vector, orthogonal to ${\cal
P}_\mu$. Hence, ``waves'' described by coordinates $Y_\mu$
(internal waves in particles, waves in the second space) are
spread with a velocity greater than the velocity of light $c$.
Thus, such a non-point particle is an original alloy of a usual
particle (variables ${\cal P}_\mu$) and a tachion (variables
${\cal Q}_\mu$).

      Observe a number of features of the suggested mechanism
for creating particle fields. First of all it may be noted that
$c$-numerical fields, resulting from the transition in the
relativistic bi-Hamiltonian system from state $f$ to state $\dot
f$, are $L$-spiral fields, since skeletons of particles $O^\Sigma
( \varphi) $ are constructed from the only $L$-spinors
$\varphi_\alpha$ and their complex conjugates
$\bbar\varphi_\alpha$ (among canonical variables of the system are
no $R$-spinors!). Thus $P$ and $C$ symmetries in the given theory
are broken right from the start. However, as will be shown in the
future, for massive particles, as a result of evolution of
$L$-fields (which will be called fundamental), there will be other
spiral fields in space-time, in particular $R$-fields according to
the first-order differential equations that satisfy particle
fields $\left(\Gamma_\mu^\Sigma\frac{\partial}{\partial
X_\mu}+M_\Sigma\right)\psi^\Sigma (X)=0$. These are not postulated
here, but are obtained as a simple consequence of symmetry
properties of space-time, in particular, of properties evolving
from Lorentz transformations of fields $\psi^\Sigma (X;Y)$
(compare with \cite{10} where similar arguments yield Dirac
equations). Next, due to theoretical assumption of one hundred per
cent fermion-antifermion asymmetry (the values $\varphi_\alpha$
can only be free, all complex conjugate ones $\bbar\varphi_\alpha$
and $\bbar\varphi$ are connected with $\varphi_\alpha$ or
$\varphi$, see \cite{1}) only the positive-frequent part (see
formula (\ref{eq43}) containing the $\theta ({\cal
P}_0)$-function) of  fermion fields appear. Negative-frequent
parts of fields as well as antifermion fields will follow from
switching to interactions, the subject that will be discussed in
the future.

\section {Problem of mass spectrum of particles}
      The mass spectrum of ``bare'' fundamental hadrons ---
baryons and mesons (particles arising from the quantum transition
$f\to\dot f$) --- is determined by formulae (\ref{eq313}) and
(\ref{eq314}).  The mass spectrum formula can be written in the
form
\begin{eqnarray}
M^2_\Sigma=2\mu^2\left(\frac{kh}{c}\right)^2\biggl\lbrace
N+5-\mu^2+\nonumber\\
+(-1)^{F+1}\sqrt{\mu^4-2\mu^2 (N+5)+Y^2+2N+9}\biggr\rbrace
\label{eq51}
\end{eqnarray}
in which the baryon (fermion) branch has $F=1$, and the meson (boson)
branch has $F=0$. Thus, in the  present theory the topic regarding
the mass spectrum of fundamental particles is connected with the
problem of finding eigenfunctions of the operator
$\stackrel{\wedge}{M^2}$ (the operator of square mass) for states
$f^\Sigma(x)$ of quanta $f$.  Moreover,  the existence of two
branches --- baryons and mesons --- is a direct consequence of the
contribution of statistical properties to the ensemble of quanta
$f$, in the form of the function $w_f\simeq n_f$, which at the
limit of strong degeneration, not being reached for above
mentioned reason, is divided into two distribution functions:  one
for fermions $n_f^F\ (F=1)$ and one for bosons $n^B_f\ (F=0)$,
described by the uniform formula
\begin{equation}
n_f=\frac{1}{\exp\left(\frac{\pi\rho}{2\mu^2}\right)+(-1)^{F+1}}.
\label{eq52}
\end{equation}
For large $\frac{\pi\rho}{2\mu^2}$ these formulae have
the same Boltzmann limit described by formula (\ref{eq214}).
The  appearance of the factor $(-1)^{F+1}$ in (\ref{eq51}) is
caused by the explicit form of function (\ref{eq52}) and follows
from the more exact dispersion law:
\begin{eqnarray}
{\cal D}_\Sigma(X)=\left(1+(-1)^Fn_f\right)\left(1+2(-1)^F
n_f\right)X^2+\nonumber\\
+4\mu^2\left(\mu^2-(N+5)(1+(-1)^Fn_f)\right)X+
\nonumber\\ +4\mu^4\left(-Y^2+(N+4)^2\right)=0
\label{eq53}
\end{eqnarray}
where $X=2\pi\rho$, and $n_f(X)$ is (\ref{eq52}),
corresponding to states $f^\Sigma=n_f O^\Sigma (\varphi)$.
Roughly speaking, for the same values of quantum numbers $N\,,\,Y$
a greater root of equation (\ref{eq53}) corresponds to
fermions\footnote{Comparing roots of equations (\ref{eq53}) or
(\ref{eq41}) with fermion and boson masses the number ${\rm
sgn}\,{\cal D}_\Sigma^{\,'} (X)$ (here $X=\frac {M^2}{4\mu^2}$)
called a dispersion curve index ${\cal D}_\Sigma (X)$ plays an
important role. The value of the index in points of crossing the
curve ${\cal D}_\Sigma (X)$ with the straight line $4\mu^2X$ is
equal to ${\rm sgn}\,{\cal
D}_\Sigma^{\,'}\left(\frac{M^2_{\Sigma\,M}}{4\mu^2} \right)=1$
(normal dispersion) and ${\rm sgn}\,{\cal D}_\Sigma^{\,'}
\left(\frac{M^2_{\Sigma\, B}}{4\mu^2}\right)=-1$ (anomalous
dispersion), i.\ e.\ ${\rm sgn}\,{\cal
D}_\Sigma^{\,'}\left(\frac{M^2_\Sigma}{4\mu^2} \right)=(-1)^F$.
This theorem (describing the connection between roots of equation
(\ref{eq311}) and fermion charges) is a consequence of an average
energy {\Large $\overline{\pi_{\mbox{\scriptsize 0}}^{\mbox
{\scriptsize\it (F)\vphantom {a}}}}$} of the fermion skeleton
$O^\Sigma$ at any temperature $T$ being greater than an average
energy {\Large $\overline {\pi_{\mbox{\scriptsize
0}}^{\mbox{\scriptsize\it (B)\vphantom{a}}}}$} which falls into
the boson skeleton. The ratio {\Large
$\overline{\pi_{\mbox {\scriptsize 0}}^{\mbox{\scriptsize\it
(F)\vphantom{a}}}}/\,\overline {\pi_{\mbox{\scriptsize
0}}^{\mbox{\scriptsize\it (B)\vphantom{a}}}}$} is a monotone
function $T$, equal to $\infty$ at $T=0$ and $\frac{9}{8}$ at
$T=\infty$ \cite{9}. In our case $T=T_f\ll 1$ (see below).}
because of the fermion skeletons being pushed aside. The
``forces'' are that help to break off the continuum on separate
points and transform it into a discontinuum.  Forces, sticking
together separate points of the discontinuum transforming it into
a continuum, are an usual interactions (strong, electromagnetic,
gravitational) between particles (see below).  Formula
(\ref{eq312}) may be obtained when $n^F_f\ll 1$.

      The appearance of leptons is connected with a transition
of the relativistic bi-Hamiltonian system in the vertex of the cone.
In this case a 4-momentum of quanta $\dot f$ is $\rho_\mu=0$
(for $C=1$ and $\dot f=1/Z$) is equivalent to $\mu^2=0$ in
formula (\ref{eq51}). Hence we obtain that the primeval masses of
all leptons $M^2_\Sigma=0$ are a trivial result which the
Twistor Program has met \cite{11} (because of that it has not
gone).

      It is meaningful to consider another limit case of formula
(\ref{eq51}) namely when $\mu^2\to\infty$. As it follows from
(\ref{eq51}), it exists only for the fermion branch ($F=1$) and
will be considered henceforth within the framework of the
realistic model $h^{(*)}_{16} $.  So now we move over to the
description of the latter model.

\section [Model of the relativistic bi-Hamiltonian system \dots]
{Model of the relativistic bi-Hamiltonian system $h^{(*)}_{16}$}
\label{kd6}
      So far we have considered  the most general basic questions
of particle field theory by a reconstruction in the model
$h^{(*)}_8$, in which isotopic symmetry is represented by the
group $U(1)$ (generator $\bar\phi\phi$),  in which for that reason
no charged particles exist.

      The realistic isotopic symmetry  described by the group
$U(2)$ (for the same Lorentz symmetry) is contained in the model
of the relativistic bi-Hamiltonian system based on the Heisenberg
algebra $h^{(*)}_{16}$. The generators, being canonical variables
of the system denoted by $\phi_{\alpha k}\,,\,\bar\phi_{\beta m}$
(here $\alpha\,,\,\beta=1,2,3,4$ are  Lorentz or Dirac indices,
and $k\,,\, m=1,2$ are isotopic) obey more general commutation
relations (\ref{eq12}).

      1) At first we briefly describe the structure of this
system given by relations (\ref{eq11}). Dynamical variables of the
system represent every possible bilinear form of canonical
variables: $\bar\phi_{\alpha k}\phi_{\beta m}\,,\,\phi_{\alpha k}
\phi_{\beta m}\,,\,\bar\phi_{\alpha k}\bar\phi_{\beta m}$. They
form a Lie algebra of dynamical variables, denoted by
$d$, that is isomorphic to the Cartan algebra $sp^{(*)}(8,{\bf C})$
of dimension 136. Real variables $\bar\phi_{\alpha k}\phi_{\beta
m}$ which conveniently written down  using of Dirac matrices
$\gamma_N$ and isotopic Pauli matrices $\tau_k$ in the form
$\bar\phi\gamma_N\tau_k\phi$ or
\begin{equation}
\eqalign{
p_\mu=i\bar\phi\gamma_\mu P_+\phi,\quad\dot p_\mu=-i\bar\phi
\gamma_\mu P_-\phi,\quad I_{\mu\nu}=\bar\phi\Sigma_{\mu\nu}\phi,\\
A=-\bar\phi\phi-4,\quad B=-\bar\phi\gamma_5\phi-4,\\
\vec\imath=\frac{1}{2}\bar\phi\vec\tau\phi,\quad\vec k=\frac{1}{2}
\bar\phi\gamma_5\vec\tau\phi,\quad Q=\frac{1}{2}\bar\phi(1+
\tau_3)\phi }
\label{eq61}
\end{equation}
as well as $\vec p^{\,\pm}_\mu=\pm i\bar\phi\gamma_\mu P_\pm\vec
\tau\phi$ and $\vec I_{\mu\nu}=\bar\phi\Sigma_{\mu\nu}\vec\tau\phi$
play the most important role. The physical sense of these
variables follows from the commutation relations which they
satisfy and is analogous to corresponding variables in the model
$h^{(*)}_8$; in (\ref{eq61}) $Q$ is the operator of electric
charge.

      The extended Fock representation is given by operators
\begin{equation}
\phi={\partial/\partial\bbar\varphi_{\alpha k}\choose
\varphi_{\alpha k}},\quad\bar\phi=\left(\bbar
\varphi_{\alpha k}\,,-\frac{\partial}{\partial
\varphi_{\alpha k}}\right),\quad\alpha, k=1,2
\label{eq62}
\end{equation}
and is constructed in the dual pair of topological vector spaces
$(\dot{\bf F},{\bf F})$ where ${\bf F}$ (and $\dot{\bf F}$)
has the already known structure ${\bf F}={\cal F}_F\otimes
{\cal F}_0$. Here ${\cal F}_F$ is a space of functions
depending on variables $\varphi_{\alpha
k}\,,\,\bbar\varphi_{\alpha k}$ (coordinates on the Lagrangian
plane $L\subset h^{(*)}_{16}$), and ${\cal F}_0$ is a space of
functions depending on additional variables (see
previous remark) $\varphi_k
\dot=\varphi_{2k}\,,\,\bbar\varphi\dot=\bbar\varphi_{2\bbar k}$
(Lorentzian scalars, i.\ e.\ scalars of the group $GL_\ell (2,{\bf
C})$ are generators $I_{\mu\nu}\,,\,A_\ell \,,\,B_\ell$)
representing isospinors (spinors of the group $GL_i(2,{\bf C})$
are generators $\vec\imath,\vec k,A_i\,,B_i$; concerning a
separation of operators $A$ and $B$ see \cite{4}). The states of
the system satisfy equations (\ref{eq21}).  But if formerly from
(\ref{eq21}) it followed the equations
$p_\mu\frac{\partial}{\partial x_\mu}f(x)=0\,,\, \dot
p_\mu\frac{\partial}{\partial\dot x_\mu}\dot f(\dot x)=0$ (as
$p^2_\mu=\dot p^2_\mu=0$), then now (as
$p_\mu=\pi^{(1)}_\mu+\pi^{(2)}_\mu$ where $\pi^{(k)}_\mu=
\bbar\varphi_k\!\!\stackrel{+}{\sigma}_\mu\!\!\varphi_k$) we have
$p^2_\mu=2\pi^{(1)}_\mu\pi^{(2)}_\mu=-4\vert\det\varphi_{\alpha k}
\vert^2=-\kappa$ (as $\sum^4_{\mu=1}\left(\pi_\mu^{(k)}\right)^2=0$),
and from (\ref{eq21}) we obtain $\left(p_\mu\frac{\partial}{\partial
x_\mu}-4i\vert\det\varphi_{\alpha k}\vert^2\right)f=0$. As the form
$p_\mu\frac{\partial}{\partial x_\mu}$ is real, and $i\vert\det
\varphi_{\alpha k}\vert^2$ is strictly imaginary, the latter equation
is equivalent to two equations
\begin{equation}
p_\mu\frac{\partial}{\partial x_\mu}f=0,\quad
\vert\det\varphi_{\alpha k}\vert^2f=0.
\label{eq63}
\end{equation}

      We cannot consider $\det\varphi_{\alpha k}=0$, since
all $\varphi_{\alpha k}$ are independent so that in the model
$h^{(*)}_{16}$, as opposed to $h^{(*)}_8$, there is no strong
relation $p_\mu^2=0$). Using  Dirac's terminology and notations
\cite{12}, we say that equations (\ref{eq63}) define a weak
coupling $\det\varphi_{\alpha k}\approx 0$ that only acts on
states $f$ or on the measures of initial states $d\mu_f$ (such
states will be called quanta $f$).

      The Lorentz-invariant measure $d\mu$ of the Lagrangian plane
$L$ is of the form
\begin{equation}
d\mu=\prod_{\alpha, k=1,2}\frac{i}{4\pi}\,d\varphi_{\alpha
k}\wedge d\bbar\varphi_{\alpha k}\,.
\label {eq64}
\end{equation}
This measure can be expressed in terms of variables $\pi^{(k)}_\mu$
by the following:
\begin{eqnarray}
d\mu=\frac{1}{16}\,\prod_{k=1,2}d\omega_k\,\delta ({\pi^{(k)}}^2)\,
\theta(\pi_0^{(k)})\,d^4\pi^{(k)}=\nonumber\\
=\theta (\pi_0)\,\theta (-\pi^2)\,\delta (\pi^2+\kappa)\,d^4\pi\,
d^4\tilde\nu\,d\kappa
\label{eq65}
\end{eqnarray}
where $\omega_k=\arg\varphi_{2k}$, and
\begin{equation}
d^4\tilde\nu=\frac{1}{16}\,d\omega_1\,d\omega_2\,
\delta (\Pi^2)\,\delta (\pi^2-2\pi\Pi)\,d^4\Pi\frac{\theta (\Pi_0)
\theta (\pi_0-\Pi_0)}{\theta (\pi_0)}
\label{eq66}
\end{equation}
(here denote $p_\mu=\pi_\mu\,,\,\pi^{(2)}_\mu=\Pi_\mu$).
According to (\ref{eq63}) the measure $d\mu_f$ represents a
contraction of the measure $d\mu$ on the ``light'' cone $\kappa=0$
and is explicitly written (the weak coupling $\det\varphi_{\alpha
k}\approx 0$)
\begin{equation}
d\mu_f=\frac{1}{(2\pi)^{3/2}}\,\theta (\pi_0)\,\delta (\pi^2)\,
d^4\pi\,d^4\nu
\label{eq67}
\end{equation}
where $d^4\nu=\left(\frac{2}{\pi}\right)^3\,d^4\tilde\nu$. The
measure $d\mu_f$ is normalized so that formula (\ref{eq211})
is valid for coherent states, thus the measure $d^4\nu$ is
normalized by the condition $\int d^4\nu=1$ (see \ref{dix2}). In
the model $h^{(*)}_{16} $ by $O^\Sigma (\pi)$ in (\ref{eq211})
we understand the integral
\begin{equation}
O^\Sigma (\pi)=\int d^4\nu\,O^\Sigma (\varphi)
\label{eq68}
\end{equation}
where $O^\Sigma (\varphi)$ are skeletons of particles in the model
$h^{(*)}_{16}$.

      The differences  between the coherent fields and their
secondary quantized Lagrangian analogues are
1) not being quantized,
2) under compression (collaps) being desintegrated into separate
Fourier-components --- non-Lagrangian fields $f^\Sigma (x)$, i.\
e.\ quanta $f$ (or partons) representing solutions of equations
(\ref{eq21}). In \cite{1} this process is referred to as the phase
transition ``particles $\Leftrightarrow$ quanta $f$''. It is
rather essential to remark that in the model $h^{(*)}_{16}$ all
coherent states are electroneutral, since integration (see
(\ref{eq68})) of charged skeletons or skeletons with hypercharge
$Y\neq 0$ such as $\bbar\varphi\vec\tau\varphi_\alpha$ or
$\varphi_\alpha$ gives zero. This conclusion is rather essential
for cosmology: before the first Big Bang (total transition
$f\to\dot f$) the Universe consisted of a mix of two
neutral gases --- bosons $X^0$ and fermions $\nu$ (see
(\ref{eq34})).

      In the model $h^{(*)}_{16}$ the solutions of equations for
$\dot f$ are written in a form similar to (\ref{eq24}):
\begin{equation}
\dot f_z=\frac{C}{Z}\,\exp (\bbar z_k\varphi_k-\bbar\varphi_k z_k)
\label{eq69}
\end{equation}
where $z_k={z_1\choose z_2}$ are complex parameters representing
an isospinor. From (\ref{eq69}) it follows that $\rho^0_\mu=-\bbar zz
(0,0,1,i)$ so that there is a weak coupling $\dot p^{\,2}_\mu\approx 0$
(in the model $h^{(*)}_8$ there was a strong coupling). Thus, as
well as in the model $h^{(*)}_8$, $\rho_\mu$ takes on values from
the lower half of light cone $N_-$~which has the invariant measure
$\frac{2} {\pi}\,\theta (-\rho_0)\,\delta
(\rho^2)\,d^4\rho=d\mu_{\dot f}$ (the measure of final states or
quanta $\dot f$).

      To consider amplitudes of  hadron transitions as an integral
over the product $\dot f_z(\dot x)f^\Sigma (x)$ with respect to
the measure of initial and final states (see (\ref{eq31}),
(\ref{eq32})), one can obtain inhomogeneous equations, similar to
(\ref{eq311}), for them and the mass spectrum formula for ``bare''
hadrons (analogue of formula (\ref{eq51}), see \ref{dix1})
\begin{eqnarray}
M^2_\Sigma=2\mu^2\left(\frac{kh}{c}\right)^2\biggl\lbrace
N+6-\mu^2+\nonumber\\
+(-1)^{F+1}\sqrt{\mu^4-2\mu^2(N+6)+4i(i+1)+2N+4}\biggr\rbrace
\label{eq610}
\end{eqnarray}
where $i$ and $N$ are spin and the isotonic quantum number of a
skeleton $O^\Sigma (\varphi)$, and $\mu^2=3T_fT_{\dot f}$ where
now $T_{\dot f}=\frac{1}{3}\bbar z_k z_k$.

      Formula (\ref{eq610}) results in a number of verified
predictions. So according to this formula the isovector
$\Sigma\,$-hyperon mass $M_\Sigma$ is more than the isoscalar
$\Lambda$-hyperon mass $M_\Lambda$ ($M_\Sigma >M_\Lambda$),
and the isovector $\rho\,$-meson mass $M_\rho$ is less than
the isoscalar $\omega$-meson mass ($M_\rho < M_\omega$) that
is in overall agreement with experimental results (see
\ref{dix4}) (for these particles $N=2$). Moreover, the ratio
$M_\Lambda M_\omega/M_\Sigma M_\rho$ depends neither on the
parameter $\mu^2$ nor on fundamental constants $c,h,k$, but is
equal to the number $\sqrt {\frac{7}{6}} \simeq$1,08. The
experimental value of this ratio is 1,16$\div$0,76 (the inaccuracy
is caused by the large width of $\rho\,$-mesons).

      In the model $h^{(*)}_{16}$, as well as in the model
$h^{(*)}_8$, the hadron fields are presented in the form of
(\ref{eq42}), (\ref{eq45}) where
\begin{equation}
O^\Sigma ({\cal P,Q})=\frac{1}{Z}\,\int d^4\nu\,O^\Sigma (\varphi)\,
e^{-\frac{-{\cal P}^2}{4\mu^2}-2i\,\Im\!\mbox{\scriptsize\it m}\,
\bbar z_k\varphi_k}\,,
\label{eq611}
\end{equation}
and $O^\Sigma (\varphi)$ are skeletons of particles in the given
model. Calculations (see \ref{dix2}) give the following
factorized expression for $O^\Sigma ({\cal P,Q})$:
\begin{equation}
O^\Sigma ({\cal P,Q})=\frac{1}{Z}\,N_\Sigma (X)\,O^{(i)}_\Sigma
(z)\,O^{(\ell)}_\Sigma({\cal P, Q}),\quad X=\sqrt{-{\cal P}^2}
\label{eq612}
\end{equation}
where $O^{(\ell)}_\Sigma$ and $O^{(i)}_\Sigma$ are Lorentzian and
isotopic wave-functions. In particular, for the baryon octet
the skeletons are
\begin{equation}
\eqalign{
O^N(\varphi)=\varphi_{\alpha k}\,,\quad O^\Lambda (\varphi)=
\varphi_{\alpha k}\,\bbar\varphi_k\,,\quad O^\Sigma (\varphi)=
\bbar\varphi\vec\tau\varphi_\alpha\,,\\
O^\Xi(\varphi)=\frac {2}{\sqrt{-{\cal P}^2}}\,(\bbar\varphi
\varphi_\alpha)\,(\bbar\varphi_k\stackrel{+}{\rho}\varphi_m)
\,\bbar\varphi_m}
\label{eq613}
\end{equation}
(the quantum number $N$ for nucleons, $\Lambda$-, $\Sigma\,$- and
$\Xi$-generators are equal to 1, 2, 2, 5 respectively),
the isotopic factors are of the form
\begin{equation}
O^{(i)}_N=z_k\,,\quad O_\Lambda^{(i)} =1\,,\quad
O^{(i)}_\Sigma=\frac{\bbar z\vec\tau z}{\bbar zz}\,,
\quad O^{(i)}_\Xi=\frac{\bbar z_k}{\bbar zz}\,.
\label{eq614}
\end{equation}
Lorentzian factors for these  particles (spin 1/2) are equal in
value, namely $O^{(\ell)}_\Sigma=\bbar\pi a$ ($a$ is a constant
spinor, see above). For the factor $N_\Sigma$ in these cases we
have
\begin{equation}
\eqalign{ N_N=\left(\frac{2}{M_N}\right)^2{\cal J}_2(M_N)\,
e^{-\frac{M^2_N}{4\mu^2}}\,,\quad N_\Lambda=\frac{2}
{M_\Lambda}\,{\cal J}_1(M_\Lambda)\,e^{-\frac
{M^2_\Lambda}{4\mu^2}}\,,\\
N_\Sigma=-\frac{2}{M_\Sigma}\,{\cal J}_3(M_\Sigma)\,
e^{-\frac{M^2_\Sigma}{4\mu^2}}\,,\quad
N_\Xi={\cal J}_2(M_\Xi)\,
e^{-\frac{M^2_\Xi}{4\mu^2}}}
\label{eq615}
\end{equation}
where ${\cal J}_n$ is the Bessel function.

      Similar expressions can be obtained for the vector
octet of mesons as well as for other multiplets.

      The factor $O^{(i)}_\Sigma\,N_\Sigma (M_\Sigma)$ contains very
important  information about the mechanism of particle creation:
it defines a priori the creation probability of a hadron $\Sigma$
in the quantum transition $f^\Sigma\to\dot f_z$ (in the
non-unitary theory by definition the state $f^\Sigma$ has a slope
in relation to $\dot f_z$, and the angle between these states
defines the given probability).  The mentioned probability
$W_\Sigma$ is defined as $\vert O^{(i)}_\Sigma \,N_\Sigma\vert^2$.
And it is clear that the total probability as a sum of partial
probabilities $W_\Sigma$ must be equal to one
\begin{equation}
\sum_\Sigma W_\Sigma=1\,.
\label{eq616}
\end{equation}
In fact, condition (\ref {eq616}) results from the normalization
condition of skeletons $O^\Sigma (\varphi)$. Due to the one
hundred per cent baryon-antibaryon asymmetry (see \cite{4}), the
skeletons are constructed (in the Lorentzian system connected with
the space ${\cal F}_0$) exclusively from additional variables
$\varphi_k$. To represent an irreducible finite-dimensional
representation basis of the group
$G_{\stackrel{\wedge}{M^2}}=GL_\ell (2,{\bf C})\otimes
U_i(2)\otimes H_i(1)$ (see paragraph \ref{kd2}; on the space
${\cal F}_0$ this group is $U_i(2)\otimes H_i(1)$, since $\varphi$
are scalars), they are written in the form \cite{13}
\begin{equation}
O^\Sigma (\varphi)=\frac{\varphi_1^m\,\varphi_2^{2i-m}}
{\sqrt{m!\,(2i-m)!}}=f_m^{(i)}
\label{eq617}
\end{equation}
where $i$ is isospin, and $m$ is its projection so that
\begin{displaymath}
\sum_{m=0}^{2i}\vert f^{(i)}_m\vert^2=\frac{\left(\vert
\varphi_1\vert^2+\vert\varphi_2\vert^2\right)^{2i}}{(2i)!}=
\frac{\left(\bbar\varphi\varphi\right)^{2i}} {(2i)!}=w_i\,.
\end{displaymath}
As can be seen, for a fixed $i$ the value $m$ satisfies the
Bernoulli distribution. The isospin $i$ satisfies the Poisson
distribution since \begin{equation}
\sum_{i=0,\frac{1}{2},1,\dots}w_i\, e^{-\bbar\varphi\varphi}=1.
\label{eq618}
\end{equation}
As $\varphi$ accepts small values (small oscillations, see below),
it is possible to put
\begin{equation}
\sum_i w_i=1.
\label{eq619}
\end{equation}

      One remark. If a bi-Hamiltonian fiber is imagined as a sea of
coupled additional variables $\bbar\varphi\varphi$,  the
configuration $\left(\bbar\varphi\varphi\right)^n$ arises in it
under (\ref{eq618}) with the probability given by the Poisson
distribution
\begin{displaymath}
P(n)=\frac{\left(\bbar\varphi\varphi\right)^n}{n!}\,e^{-\bbar
\varphi\varphi}\,.
\end{displaymath}
It is natural to consider functions $u^{(i)}_m=\exp\left(-\frac{\bbar
\varphi\varphi}{2}\right)\,f^{(i)}_m$ forming an orthonormal
system  of functions in the space ${\cal F}_0$ in relation with
the scalar product
\begin{displaymath}
(f,g)=\int_{{\bf C}^2}\overline{f(\varphi)}\,g(\varphi)\,
d\mu(\varphi)
\end{displaymath}
with measure
\begin{displaymath}
d\mu(\varphi)=\prod_{k=1,2}\frac{i}{4\pi}\,d\varphi_k\wedge
d\bbar\varphi_k
\end{displaymath}
(in fact $\left (u^{(i)}_m\,,\,u^{(i')}_{m'}\right)=\delta_{ii'}
\,\delta_{mm'}$)
for which the condition
\begin{displaymath}
\sum_{i=0,\frac{1}{2},1,\dots}\sum_{m=0}^{2i}\vert u^{(i)}_m
\vert^2=\sum_ie^{-\bbar\varphi\varphi}\,w_i=1
\end{displaymath}
is satisfied.

      Let us return to condition (\ref{eq619}) which took
place before the transition $f\to\dot f$. After the transition
$f\to\dot f$ and the creation of fundamental hadron fields this
normalization condition of skeletons passes to condition
(\ref{eq616}).

      We now show that in (\ref{eq618}) the exponent can be put
equal to 1, so $\vert\varphi\vert$ is a small value. Both the
Bernoulli and Poisson distribution are realized in each fiber.
Another distribution used by us --- the Gibbs distribution
$\exp\left(\frac{-\bbar\varphi\varphi}{T_f}\right)$ --- describes
the statistics of fibers. It follows from this that values
$\bbar\varphi\varphi\sim T_f$ as well as $\vert\varphi
\vert\sim\sqrt{T_f}$. We now see that $T_f\sim 10^{-6}$ so
that the value $\vert\varphi\vert\sim 10^{-3}$ is really small.

      Dimensionless parameters of the theory, such as
$T_f\,,\,z_k$ (or the sum $T_{\dot f}=\frac{1}{3}\left(\vert z_1
\vert^2+\vert z_2\vert^2\right)$ and ratio $\varepsilon =\vert z_1/z_2
\vert$) and $1/Z$, are necessary to perform calculations in the framework
of the present theory. The factor $1/Z$ was calculated in \cite{7}.
Here parameters $T_f\,,\,T_{\dot f}$ and $\varepsilon$ will be found.

      At first we define the parameter $\mu^2=3T_fT_{\dot f}$.
It is used in formula (\ref{eq610}) defining the mass spectrum of
``bare''(non-interacting) hadrons and is found from conditions
(generally named the minimum principle) put on factors of the
quadratic form $P_\Sigma (X)=\left(X-M^2_{\Sigma\,B}
\right)\left(X-M^2_{\Sigma\,M}\right)$. Denote by $P_{\Sigma_0}
(X)$ the form for which a baryon root $M_{\Sigma\,B}$ has the
least value. This value corresponds to the point
$i_0=-\frac{1}{2}$ and $N_0=-1$ and is equal to
$M^2_{\Sigma_0\,B}=2\mu^2\left\{5-\mu^2+\sqrt
{\mu^4-10\mu^2+1}\right\}$. The form $P_{\Sigma_0}(X)$ accepts
the least value at $X=\frac{1}{2}\left(M^2_{\Sigma_0\,B}+
M^2_{\Sigma_0\,M}\right)$, equal to
$-\frac{1}{4}\left(M^2_{\Sigma_0
\,B}-M^2_{\Sigma_0\,M}\right)^2=-4\mu^4\left (\mu^4-10\mu^2+1\right)$.
The latter expression as a function of the parameter $\mu^2$ has
a minimum at
$\mu^2=\frac{15}{4}\left(1-\sqrt{1-\frac{8}{225}}\right)
\simeq$0,067 (determined by the equation $2\mu^4-15
\mu^2+1=0$).

      The other parameter $\bbar z_k z_k$ is used in condition
(\ref{eq616}) in the same way as the parameter $\mu^2$ . Since
proton and neutron take on the greatest probability in the
sum (\ref{eq616}) (herein it is not difficult to be convinced
looking at formulae (\ref{eq614}), (\ref{eq615})), condition (\ref
{eq616}) with large accuracy is written in the form
\begin{displaymath}
\bbar zz\left(\frac{2}{M_N}\right)^4{\cal J}^2_2(M_N)\,
e^{-\frac {M^2_N}{2\mu^2}}\simeq 1\,.
\end{displaymath}
If $M_N$ takes the value $M_N=$1,14 given by formula
(\ref{eq610}) at $\mu^2=$0,067 ($N=1,\,i=\frac{1}{2}$), for
$\bbar zz$ we obtain $\bbar zz=$0,86$\cdot 10^5$. Thus $T_{\dot
f}=$0,3 $\cdot 10^5$. In the same time for
$T_f=\frac{\mu^2}{3T_{\dot f}}$ we have $T_f=$0,78$\cdot 10^{-6}$.
Another dimensionless parameter $\eta=\frac{3T_{\dot f}}{T_f}$,
playing an important role in cosmology, is equal to
$\eta=10^{11}$. The parameter $\varepsilon$ will be determined
below.

      2) Representations of algebra (\ref{eq12}). In this case
the coordinates on the Lagrangian plane $L\subset h^{(*)}_{16}$ are
still denoted by $\varphi_{\alpha k}\,\,\bbar\varphi_{\alpha k}$.
The representation of algebra (\ref{eq12}) given by operators
\begin{equation}
\phi^H={\Lambda\partial/\partial\bbar\varphi_{\alpha k}\choose
\varphi_{\alpha k}},\quad\bar\phi^H=\left(\bbar
\varphi_{\alpha k}\,,-\Lambda\frac{\partial}{\partial
\varphi_{\alpha k}}\right)
\label{eq620}
\end{equation}
is called the $H$-representation (or hadronic). In this
representation dynamical variables of the system (quadratic
Hamiltonians) form the algebra $d=\{d_-,d_0,d_+\}$ graduated
by powers of $\Lambda$ in which
\begin{eqnarray}
d_+\,:\qquad\Lambda\,\partial_{\alpha k}\,\partial_{\beta m}\,,
\quad\Lambda\,\bar\partial_{\alpha k}\,\bar\partial_{\beta m}\,,
\quad\Lambda\,\partial_ {\alpha k} \,\bar\partial_ {\beta m}
\nonumber\\
d_-\,:\quad\frac{1}{\Lambda}\,\varphi_{\alpha k}\,
\varphi_{\beta m}
\,,\quad\frac{1}{\Lambda}\,\bbar\varphi_{\alpha
k}\,\bbar\varphi_{\beta
m}\,,\quad\frac{1}{\Lambda}\,\bbar\varphi_{\alpha
k}\,\varphi_{\beta m} \nonumber\\
d_0\,:\qquad\varphi_{\alpha
k}\,\partial_{\beta m}\,,\ \bbar\varphi_{\alpha
k}\,\bar\partial_{\beta m}\,,\ \bbar\varphi_{\alpha
k}\,\partial_{\beta m}\,,\ \varphi_{\alpha k}\,\bar
\partial_{\beta m}\nonumber
\end{eqnarray}
(here $\partial_{\alpha k}=\frac{\partial}{\partial
\varphi_{\alpha k}} \,,\,\bar\partial_{\beta
m}=\frac{\partial}{\partial\bbar \varphi_{\beta m}}$).
Obviously we have
\begin{displaymath}
[d_0\,,\,d_0]\subset d_0\,,\quad
[d_\pm\,,\, d_0]\subset d_\pm\,, \quad
[d_+\,,\,d_+]=[d_-\,,\,d_-]=0,\quad
[d_+\,,\, d_-] \subset d_0\,.
\end{displaymath}

      Another  representation, the compressed one (or long-drawn
one, if $\Lambda >1$) in relation to (\ref{eq620}) given by
operators
\begin{equation}
\phi^L={\partial/\partial\bbar\varphi_{\alpha k}'\choose
\Lambda\varphi_{\alpha k}'},\quad\bar\phi^L=\left(\Lambda\bbar
\varphi_{\alpha k}'\,,-\frac{\partial}{\partial
\varphi_{\alpha k}'}\right),
\label{eq621}
\end{equation}
where $\varphi_{\alpha k}'$ is connected with $\varphi_{\alpha k}
$ by a (generally speaking) purely isotopic transformation $V\,:\
\varphi_{\alpha k}'=V_{km}\,\varphi_{\alpha m}$, ($V\in
SL_i(2,{\bf C})$), is called the $L$-representation (or leptonic
one).

      Momentum variables in the $H$-representation are written in the
form $^H\!p_\mu=\frac{1}{\Lambda}\,\bbar\varphi\stackrel{+}{\sigma}_\mu
\varphi=\frac{1}{\Lambda}\,\pi_\mu\,,\,^H\!\dot p_\mu=-\Lambda\,\partial
\stackrel{-}{\sigma}_\mu\bar\partial$, and in the $L$-representation
they are in the form $^L\!p_\mu=\Lambda\,\bbar\varphi\stackrel{+}{\sigma}
\varphi\,,\,^L\!\dot p_\mu=-\frac{1}{\Lambda}\,\partial\stackrel
{-}{\sigma}_\mu\bar\partial$.

      Obviously,  at $\Lambda\neq 1$ the mass spectrum formula for
hadrons remains the same as (\ref{eq610}), leaving also the
definition of 4-momenta ${\cal P}_\mu$ and ${\cal Q}_\mu$ as well
as the formula for particle fields and transition amplitudes
unchanged. However, by $X_\mu$ and $Y_\mu$ now it is necessary to
understand that
\begin{equation}
X_\mu=\frac{1}{2}\left(\frac{1}{\Lambda}x_\mu+\Lambda\dot x_\mu\right),
\quad Y_\mu=\frac{1}{2}\left(\frac{1}{\Lambda}x_\mu-\Lambda\dot x_\mu
\right).
\label{eq622}
\end{equation}

      In Appendix~3 it will be shown that in (\ref{eq12}) $\Lambda$
can impossibly be a number; $\Lambda$ turns out to be connected with
the dimension $\dim d$ of the algebra of dynamical variables $d$ by the
formula $\Lambda=\sqrt{\dim d}$. In the model $h^{(*)}_{16}$ the
dimension $\dim d=$136.

      3) Hadron and lepton era. In the present theory the creation
of particles  occurs in strict order: at first hadrons are
generated, and subsequently when the density of quanta $f$
decreases, leptons will be generated.  The creation of hadrons
(hadron era) is described by the $H$-representation of algebra
(\ref{eq12}). In this representation a 4-momentum of quanta $\dot
f$ is large $^H\!\dot p_\mu=-\Lambda\,\partial\!\stackrel
{-}{\sigma}_\mu\!\bar\partial$ (factor $\Lambda$), the state $\dot f_z$
is of the form (\ref{eq69})  where $z_k=z{\varepsilon\choose 1}$
(here $\varepsilon=z_1/z_2\,,\,z=z_2$), and a 4-momentum of quanta
$f$ is small $^H\!
p_\mu=\frac{1}{\Lambda}\,\bbar\varphi\!\stackrel{+}
{\sigma}_\mu\!\varphi$  (factor $1/\Lambda$). Therefore the
distribution function  $w_f$ (\ref{eq214}) is rather essential. As
a consequence  of the  irreversible (that aspect being
emphasized!) quantum transition $f\to\dot f$ both neutral and
charged hadrons are created. It is important to notice that
$^H\!\dot p_\mu$ and $^H\!p_\mu$ are neutral operators in the
sense  that they commute with the operator of electromagnetic
charge $Q\,:\ [^H\!\dot p_\mu\,,\,Q]=[^H\!p_\mu\,,\,Q]=0$. However
states $\dot f_z$ (at $\varepsilon\neq 0$) are characterized by an
uncertain charge not equal to zero $Q$, since $Q\,\dot f_z\neq 0$
(that is why charged hadrons occur). Due to the assumption of
complete (one hundred per cent)  fermion-antifermion asymmetry
(see \cite{4}) only the fermions will be created. The antifermion
creation will happen subsequently as a result of switching to
interactions. Clearly, in the hadron era a large positive charge,
not compensated by anything, arises (before the transition the
charge of quanta $f$ was equal to zero, see above, in the model
$h^{(*)}_{16}$ the coherent fields (\ref{eq211}) as a matter of
fact are neutral leptons with zero mass), due to the proton
component.  In fact, probabilities of creating various charged
components of the baryon octet (main hadron component) satisfy the
condition (they result from (\ref{eq615})) \begin{equation}
W_{p_+}\gg W_{\Sigma^\pm}\gg W_{\Xi^-} \label{eq623}
\end{equation} so that negative $\Xi^-$-baryons cannot compensate
the positive charge of protons.  Hence, the process creating only
hadrons would be accompanied by a violation of the electric charge
conservation law (for mesons this problem is not present: the
charge $\rho^+$ is compensated by the charge $\rho_-$, and the
charge $K^{*\,+}$ is by the charge $K^{*\,-}$). To compensate the
charge of protons only opposite charged leptons can occur due to
the following.

      After a creation of  hadrons (the transition $N_+\to\ N_-$)
the density of quanta $f$ sharply decreases, so that the density
of quanta $\dot f$ on the lower half of light cone ($\rho\in\
N_-$) increases. Now basically there are transitions in the vertex
of cone $N_-\to\{0\}$ (lepton era). This era is characterized by
the density function $w_f=1$ (as $\rho_\mu=0$ that is equivalent
to the parameter $z=0$) and is described by the (long-drawn)
$L$-representation of algebra (\ref{eq12}). In this representation
the 4-momentum $^L\!p_\mu=
\Lambda\,\bbar\varphi'\stackrel{+}{\sigma}_\mu\varphi'$ is large (the
factor is $\Lambda$), and $^L\!\dot
p_\mu=\frac{1}{\Lambda}\,\partial\,'
\!\stackrel{-}{\sigma}_\mu\!\bar\partial\,'$ is small (the factor is
$\frac{1}{\Lambda}$).  Graphically speaking, the yoke ($^H\!\dot
p,\, ^H\!P$) can overturn, as a result it can take another
position ($^L\!\dot p,\,^L\!p$) such that when $p$ rises, $\dot p$
drops to slide in $\{0\}$ (in the vertex of cone), an additional
purely isotopic rotation $\varphi_k\to V_{km}\,\varphi_m$ occurs
(the process of transition is described by Lorentz-invariant
amplitudes, and it occurs in the space ${\cal F}_0$ of additional
variables --- Lorentzian scalars $\varphi\dot
=\varphi_2\,,\,\bbar\varphi\dot =\bbar\varphi_2$ where Lorentz
transformations are trivial). As a consequence a 4-momentum
$\frac{1}{\Lambda} \,\bbar\varphi\stackrel{+}{\sigma}_\mu\varphi$
passes into a  4-momentum $\Lambda\pi_\mu'$:
\begin{displaymath}
\frac{1}{\Lambda}\,\bbar\varphi\stackrel{+}{\sigma}_\mu\varphi\to
\Lambda\pi_\mu'=\Lambda\,\bbar\varphi'\stackrel{+}{\sigma}_\mu
\varphi'=\Lambda\,\bbar\varphi\stackrel{+}{\sigma}_\mu
V^+V\,\varphi=\frac{1}{\Lambda}\,\bbar\varphi\stackrel{+}
{\sigma}_\mu\tilde V^+\tilde V\,\varphi
\end{displaymath}
where $\varphi_{\alpha k}'=V_{km}\,\varphi_{\alpha m}\,,\,V\in SL_i
(2,{\bf C})\ (\det V=1)$, and $\tilde V=\Lambda V\in GL_i(2,{\bf C})\
(\det\tilde V=\Lambda^2)$. Since it may be taken $V^+V=\stackrel
{+}{\tau}_m\!A_m$ where $A_m$ is a time-like 4-isovector ($A^2_m=1$,
since $\det V=1$), it may be written $\pi_\mu'=\pi_\mu^A$ where
$\pi_\mu^A=\bbar\varphi\stackrel{+}{\sigma}_\mu\stackrel{+}{A}\varphi$,
and $\stackrel{+}{A}=\stackrel{+}{\tau}_\mu\!A_m$. Being distinct from
$\pi_\mu$ the 4-momentum $\pi^A_\mu$ carries an electrical charge, since
it does not commute with the charge operator $Q\,:\ [\pi^A_\mu\,,\,Q]
\neq 0$. However the state $\dot f_0=1/Z$ corresponding to the vertex
of cone is neutral: $Q\dot f_0=0$ (that is why finishing transitions
will be in the vertex of cone and that is why an isotopic rotation
$V$ becomes possible; at $z\neq 0$ the transformation is not admissible,
in this case only dilatations $\varphi\to\Lambda\varphi$ are allowed).
With the transformed 4-momentum the lepton fields are written
in the form (compare with (\ref{eq211}))
\begin{equation}
\psi^\Sigma (X+Y)=\frac{1}{(2\pi)^{3/2}}\,\frac{1}{Z}\,\int
e^{i\Lambda\pi'(X+Y)}\,\theta (\pi_0)\,\delta (\pi^2)\,d^4\pi\,
O^\Sigma (\pi)
\label{eq624}
\end{equation}
where as always
\begin{displaymath}
O^\Sigma (\pi)=\int d^4\nu\,O^\Sigma (\varphi).
\end{displaymath}
In (\ref{eq624}) it is convenient to switch from variables $\varphi$
to variables $\varphi'=V\varphi$. Since for such transformations
the measure $d\mu_f$ does not change, we come to the formula
\begin{displaymath}
\psi^\Sigma (X+Y)=\frac{1}{(2\pi)^{3/2}}
\,\frac{1}{Z}\,\int e^{i\Lambda\pi (X+Y)}\,\theta (\pi_0)
\,\delta (\pi^2)\,d^4\pi\,\tilde O^\Sigma (\pi)
\end{displaymath}
where
\begin{equation}
\tilde O^\Sigma (\pi)=\int
d^4\nu\,O^\Sigma (V^{-1} \varphi).
\label{eq625}
\end{equation}
As always, all skeletons with non-zero hypercharge and isospin will
result in zero. However, now also charged fields occur.

      For an example of the lowest states of fermions, the skeletons
are in fact of the form
\begin{displaymath}
O^\Sigma (\varphi)=\bbar\varphi\tau_n\varphi_\alpha\,,
\end{displaymath}
and the integral
\begin{displaymath}
O^\Sigma (\pi)=\int d^4\nu\,\bbar\varphi\tau_n\varphi_\alpha=
\delta_{n0}\varphi_\alpha (\pi)
\end{displaymath}
is only different from zero at $n=0$, i.\ e.\ for $\bbar\varphi
\varphi_\alpha$. This neutral lepton field is a neutrino. Next,
as $O^\Sigma (V^{-1}\varphi)=\bbar\varphi {V^+}^{-1}\tau_nV^{-1}
\varphi_\alpha$, and ${V^+}^{-1}\tau_nV^{-1}=L_{nn'}(V^{-1})
\tau_{n'}$, we have
\begin{displaymath}
\tilde O^\Sigma (\pi)=L_{nn'}(V^{-1})\,\int d^4\nu\,\bbar\varphi
\tau_{n'}\varphi_\alpha=L_{n0}(V^{-1})\varphi_\alpha (\pi)
\end{displaymath}
where $L_{n0}=\frac{1}{2}\,Sp\,{V^+}^{-1}\tau_nV^{-1}$. Fields with
$L_{00}$  and $L_{30}$ are neutral, and fields with
$L_{+0}=L_{10}+iL_{20}$ and  with
$L_{-0}=L_{10}-iL_{20}\sim\varepsilon$ are charged. As can be
seen, charged leptons occur.

      It  remains to find a special kind of the transformation
$V$. Since the given transformation is only performed with one
purpose, namely to compensate any charge, neutral fields (nonzero
before the transformation) remain so after the transformation.
In  particular, this concerns a field with the skeleton
$\bbar\varphi\tau_3 \varphi_\alpha$ which before the
transformation was equal to zero.  After the transformation it is
$\sim L_{30}$. Therefore it must be $L_{30}=0$. This means that
\begin{equation}
Sp\,{V^+}^{-1}\tau_3 V^{-1}=0.
\label{eq626}
\end{equation}
If after the transformation $\varphi\to\tilde V\varphi$ the
hadrons begined to appear, all of them should be neutral. This
means that it must be
\begin{displaymath}
\bbar Z\tilde V=\bbar Z^0
\end{displaymath}
where $Z^0\sim{0\choose 1}$ is a neutral
isospinor. From the condition
\begin{equation}
\bbar z\tilde
V\tilde V^+z=\bbar zz=\vert z\vert^2\left(
1+\vert\varepsilon\vert^2\right)
\label{eq627}
\end{equation}
(since  for the transformation $\tilde V$ neither energy nor
temperature of quanta $\dot f$ should be changed) it follows that
$z^0=z\sqrt{1+ \vert\varepsilon\vert^2}{0\choose 1}$. From
equations (\ref{eq626}), (\ref{eq627}) we find
\begin{equation}
V=\left(\matrix{\frac{\Lambda}{\sqrt{1+\vert\varepsilon\vert^2}}&
\frac{\varepsilon}{\Lambda\sqrt{1+\vert\varepsilon\vert^2}}\Biggl(1+
\frac{\sqrt{\Lambda^4-1}}{\vert\varepsilon\vert}\Biggr)\cr
-\frac{\Lambda\bbar\varepsilon}{\sqrt{1+\vert\varepsilon\vert^2}}&
\frac{1}{\Lambda\sqrt{1+\vert\varepsilon\vert^2}}\Biggl(1-
\vert\varepsilon\vert\sqrt{\Lambda^4-1}\Biggr)\cr}\right).
\label{eq628}
\end{equation}
At  last the most important condition should be thought over. If
the transformation $\tilde V$ is performed again, it may not lead
to a new position of the yoke $(\dot p,p)$ or to a new generation
of particles. Actually the system has to return to the
$H$-representation. With reference to $V$ this means that the
twice applied transformation $V$ must be proportional to an unit
transformation: $V^2=\alpha$. Hence we obtain $\det V=\pm \alpha$.
As $\det V=1$, $\alpha=\pm 1$. For matrices in the form
(\ref{eq628}) only the condition $\alpha=-1$ is acceptable,
therefore
\begin{equation}
V^2=-1.
\label{eq629}
\end{equation}

      From (\ref{eq629}) it follows that in (\ref{eq628})
\begin{equation}
\vert\varepsilon\vert=\sqrt{\frac{\Lambda^2+1}{\Lambda^2-1}},
\label{eq630}
\end{equation}
and consequently the matrix $V$ must be of the form
\begin{equation}
V=\left(\matrix{\sqrt{\frac{\Lambda^2-1}{2}}&
\sqrt{\frac{\Lambda^2+1}{2}}\,e^{i\chi}\cr
-\sqrt{\frac{\Lambda^2+1}{2}}\,e^{-i\chi}&
-\sqrt{\frac{\Lambda^2+1}{2}}\cr}\right)=
\sqrt{\frac{\Lambda^2-1}{2}}\left(\matrix{1&
\varepsilon\cr -\bbar\varepsilon & -1}\right),
\label{eq631}
\end{equation}
where $\chi=\arg\varepsilon$. Since in the model $h^{(*)}_{16}\
\Lambda^2=136$, $\vert\varepsilon\vert=\sqrt{137/135}$, and
the matrix $V$ takes the form
\begin{displaymath}
V=\frac{1}{\sqrt{2}}\left(\matrix{\sqrt{135}&
\sqrt{137}\,e^{i\chi}\cr -\sqrt{137}\,e^{-i\chi}&
-\sqrt{135}\cr}\right).
\end{displaymath}

      It is interesting to observe that a priori
probabilities of creating proton $W_p$ and neutron $W_n$
are not equal (see (\ref{eq614}))
\begin{equation}
\frac{W_p}{W_n}=\vert\varepsilon\vert^2=\frac{137}{135}=1,015\,.
\label{eq632}
\end{equation}
Thus, the  transition $f\to\dot f$ will generate slightly more
protons than neutrons.

      From the formula $V^+V=\stackrel{+}{\tau}_mA_m$ it is
possible to find components of the isovector $A_m$
\begin{equation}
A_0=\Lambda^2,\quad A_3=0,\quad A_1=\sqrt{\Lambda^4-1}\,\cos
\chi,\quad A_2=\sqrt{\Lambda^4-1}\,\sin\chi,
\label{eq633}
\end{equation}
and from  the formula $\frac{1}{2}\,Sp\,{V^+}^{-1}\tau_nV^{-1}=
L_{n0}$ the values $L_{n0}$ are
\begin{equation}
L_{00}=\Lambda^2,\quad L_{30}=0,\quad L_{+0}=-\sqrt{\Lambda^4-1}\,
e^{-i\chi},\quad L_{-0}=-\sqrt{\Lambda^4-1}\,e^{i\chi}.
\label{eq634}
\end{equation}

      4) So, hadrons arise from the transition $^H\!p\to$$^H\!\dot p$,
leptons  arise from the transition $^L\!p\to$$^L\!\dot p$. There
are two more transition possibilities: $^H\!p\to$$^L\!\dot p$ and
$^L\!p\to$ $^H\!\dot p$.

      In the first case both 4-momenta $^H\!p$ and $^L\!\dot p$ are
small (zero  as a matter of fact) so that in this case the system
passes from the vertex of the upper half of light cone into the
vertex of the lower half of light cone. This is a vacuum
transition. Its amplitude is equal to the factor $1/Z$.

      In the second case both 4-momenta are large: $^L\!p=\Lambda\,
\bbar\varphi\stackrel{+}{\sigma}_\mu\varphi\,,\,^H\!\dot p=-\Lambda\,
\partial\stackrel{-}{\sigma}_\mu\bar\partial$ ($\Lambda$ is a factor,
the matrix $V=1$). This is that limiting case when the parameter
$\mu^2\to\infty$. In this limit, being only meaningful for fermions
($F=1$), the equation (\ref{eq311}) passes into the equation (it is
necessary to pay attention on the factor $\Lambda^2$)
\begin{displaymath}
\left(\Box_X-\Lambda^2F^0_\Sigma\right)\,O^\Sigma (X;Y)=
\Im^\Sigma(X;Y),
\end{displaymath}
and formula (\ref{eq610}) gives
\begin{equation}
M^2_{T\,\Sigma}=-\Lambda^2\left[(N+5)^2+7-4i(i+1)\right].
\label{eq635}
\end{equation}
As can be seen, in our spacetime (coordinates $X_\mu$) the objects
arising  in this case are characterized by an imaginary mass. In
addition they have no wave function, since solutions of the
equations
\begin{displaymath}
\left(\Box_X-M^2_{T\,\Sigma}\right)\,\psi^\Sigma (X;Y)=0
\end{displaymath}
do not represent any fields.
Such  objects do not exist as particles. However in the second
space (coordinates $Y_\mu$) these objects represent usual
particles, since $\psi^\Sigma (X;Y)$ as functions $Y$ satisfy the
Klein-Gordon equation
\begin{displaymath}
\left(\Box_Y+M^2_{T\,\Sigma}\right)\,\psi^\Sigma (X;Y)=0
\end{displaymath}
and are characterized by real masses
\begin{displaymath}
M^2_{T\,\Sigma}=\Lambda^2\left[(N+5)^2+7-4i(i+1)\right].
\end{displaymath}
But the second (internal) space is inaccessible for observations.
In our space the objects in question can show themselves only as
virtons, being objects which are described by a distribution
function (for the lack of any wave function), compare with
\cite{14}. To go out (because of their large mass for an instant)
from the second space in our space, they  certainly have to return
into the second space where they exist as usual particles. In that
way these objects stick together different points of the
discontinuum to transform it in a continuum \cite{7}.

\section{At last}
    God gave the people a mind to  comprehend the truth, but He
did not give them the criteria of the truth, He preserved it for
Himself. That is why people often consider lies as the truth and
vice versa, and that is why we depend from God and must not lose
contact with Him.

\appendix
\section{}
\label{dix1}
      The determination of the mass spectrum of ``bare''
fundamental hadrons is connected with the determination of
eigenvalues of the operator $\stackrel{\wedge}{M^2}=2\dot p_\mu
p_\mu$ on states $f^\Sigma=w\, O^\Sigma$ of relativistic
bi-Hamiltonian system: $\stackrel {\wedge}{M^2}f^\Sigma=F_\Sigma
(-{\cal P}^2)f^\Sigma$. Here we find the expression for $F_\Sigma
(-{\cal P}^2)$.

      At first we consider the model $h^{(*)}_8$. Here 4-momenta
of the system are of the form $p_\mu=i\bar\phi\gamma_\mu P_+\phi=\bbar
\varphi_{\dot\beta}(\stackrel{+}{\sigma}_\mu)^{\dot\beta\alpha}
\varphi_\alpha\,,\,\dot p_\mu=-i\bar\phi\gamma_\mu P_-\varphi=
-\partial^\beta (\stackrel{-}{\sigma}_\mu)_{\beta\dot\alpha}
\partial^{\dot\alpha}$ where $\partial^\alpha=\frac{\partial}
{\partial\varphi_\alpha}$~. Therefore $2\dot p_\mu p_\mu=-2\partial
\stackrel{-}{\sigma}_\mu\bar\partial\ \bbar\varphi\stackrel{+}
{\sigma}_\mu\varphi=-(\stackrel{\wedge}{N}+4)^2+\stackrel{\wedge}
{Y^2}$ (we have used the completeness condition of matrices
$\sigma_\mu$: $\sum_{\mu=1}^4\left(\stackrel{+}{\sigma}_\mu
\right)^{\dot\alpha\beta}\left(\stackrel{-}{\sigma}_\mu
\right)_{\gamma\dot\delta}=2\delta^{\dot\delta}_{\dot\alpha}\,
\delta^\beta_\gamma$ and the definition $\stackrel{\wedge}{N}=
\varphi_\alpha\partial^\alpha+\bbar\varphi_{\dot\alpha}\bar
\partial^{\dot\alpha}\,,\,\stackrel{\wedge}{Y}=\varphi_\alpha
\partial^\alpha-\bbar\varphi_{\dot\alpha}\bar\partial^{\dot\alpha}$).
Since skeletons $O^\Sigma$ are eigenvectors of operators $\stackrel
{\wedge}{N}$ and $\stackrel{\wedge}{Y}$: $\stackrel{\wedge}{N}O^\Sigma=
N^\Sigma O^\Sigma\,,\,\stackrel{\wedge}{Y}O^\Sigma=Y^\Sigma O^\Sigma$
($N^\Sigma$ and $Y^\Sigma$ are eigenvalues, here and in the sequel the
index $\Sigma$ in eigenvalues is suppressed, and $w=\exp (-\omega^2
\bbar\varphi\varphi)$ is a distribution function of quanta $f$ on
energy $\varepsilon_f=\bbar\varphi\varphi$ (for a while a temperature of
quanta $f$ is denoted as $T_f=1/\omega^2$) as well as $\stackrel
{\wedge}{N}$ and $\stackrel{\wedge}{Y}$ are first-order differential
operators, $\stackrel{\wedge}{N}(O^\Sigma w)=(\stackrel{\wedge}
{N}O^\Sigma)w+O^\Sigma(\stackrel{\wedge}{N}w)$, such that by
definition $w$ it follows $\stackrel{\wedge}{Y}w=0$. Therefore we
have
\begin{displaymath}
\stackrel{\wedge}{M^2}f^\Sigma=\left[-(N+4)^2+Y^2\right]
f^\Sigma-O^\Sigma (\stackrel{\wedge}{N^2}w)-8O^\Sigma
(\stackrel{\wedge}{N}w)-2NO^\Sigma(\stackrel{\wedge}{N}w).
\end{displaymath}
Next, since
\begin{displaymath}
\stackrel{\wedge}{N}w=(\varphi\partial +\bbar\varphi\bar\partial)
e^{-\omega^2\bbar\varphi\varphi}=2\omega^2\frac{\partial}
{\partial\omega^2}w,
\end{displaymath}
it is possible to write
\begin{displaymath}
\stackrel{\wedge}{M^2}f^\Sigma=F^0_\Sigma f^\Sigma-4\left[(
\omega^2)^2\frac{\partial^2}{(\partial\omega^2)^2}+(N+5)\omega^2
\frac{\partial}{\partial\omega^2}\right] f^\Sigma
\end{displaymath}
where $F^0_\Sigma=-(N+4)^2+Y^2$ (see (\ref{eq210})). Let us now make
use of the identity $\frac {\partial} {\partial\omega^2}w=-\bbar\varphi
\varphi w$. We write $\omega^2\frac{\partial}{\partial\omega^2}w=
-\omega^2\bbar\varphi\varphi w=\frac{{\cal P}^2}{4\mu^2}w$. The latter
equality  follows from the possibility to write the Gibbs function
representation $\exp (-\omega^2\bbar\varphi\varphi)$ in the
relativistic invariant Juttner form $\exp (-\frac{-{\cal
P}^2}{4\mu^2})$ where $-{\cal P}^2=
4\pi\rho=4\bbar\varphi\varphi\bbar zz$, and $\mu^2=\bbar zz/\omega^2=
3T_fT_{\dot f}$ (see formula (\ref{eq214})). Therefore $F_\Sigma$
is written in the form
\begin{displaymath}
F_\Sigma (-{\cal P}^2)=F^0_\Sigma
+\frac{N+5}{\mu^2} (-{\cal P}^2)-\frac{(-{\cal P}^2)^2}{4\mu^4}\,.
\end{displaymath}
In the coordinate representation we have
\begin{equation}
F_\Sigma (\Box)=F^0_\Sigma +\frac{N+5}{\mu^2}
\Box -\frac{\Box^2}{4\mu^4}
\label{pr11}
\end{equation}
where $\Box$ is the D'Alembertian with respect to variables
$X_\mu$.

      In the model $h^{(*)}_{16}$ the operator $\stackrel{\wedge}
{M^2}$ is other, as here \mbox{$p_\mu=\bbar\varphi_{\dot\alpha k}
(\stackrel{+}{\sigma}_\mu)^{\dot\alpha\beta}\varphi_{\beta k}\,,$}
$\dot p_\mu=-\partial^\beta_k(\stackrel{-}{\sigma}_\mu)_{\beta\dot
\alpha}\bar\partial^{\dot\alpha}_k\ (\partial_k^\alpha=\frac
{\partial}{\partial\varphi_{\alpha k}})$. Using the completeness
condition of matrices $\sigma_\mu$~, we write
\begin{displaymath}
\stackrel{\wedge}{M^2}=2\dot p_\mu p_\mu=-4\partial^\alpha_k\,
\bar\partial^{\dot\beta}_k\,\bbar\varphi_{\dot\beta m}\,
\varphi_{\alpha m}\,
\end{displaymath}
and using the completeness condition of matrices $\stackrel{\pm}
{\tau}_a=(\vec\tau,\pm 1):\sum_a\left(\stackrel{+}{\tau}_a
\right)_{km}\left(\stackrel{-}{\tau}_a\right)_{np}$\quad$=2\delta_{kp}
\delta_{nm}\,$, we shall have (here $\tau^T$ is a transposed matrix)
\begin{displaymath}
\stackrel{\wedge}{M^2}=-2\partial\stackrel{+}{\tau}_a\varphi\
\bar\partial\stackrel{-}{\tau}_a^T\bbar\varphi=-2\partial
\stackrel{+}{\tau}_a\varphi\left(\bbar\varphi\stackrel{-}{\tau}_a
\bar\partial+4\delta_{a0}\right)=
\end{displaymath}
\begin{displaymath}
=-2\left[\partial\vec\tau\varphi\bbar\varphi\vec\tau\bar
\partial+(\partial\varphi)(\bbar\varphi\bar\partial)+
4\partial\varphi\right].
\end{displaymath}
Using then definitions of the operators $\stackrel{\wedge}
{N}=\varphi\partial+\bbar\varphi\bar\partial\,,\,\stackrel{\wedge}
{Y}=\varphi\partial-\bbar\varphi\bar\partial\,,\,\stackrel
{\wedge}{\vec\imath}=-\frac{1}{2}(\varphi\vec\tau^T\partial-
\bbar\varphi\vec\tau\bar\partial),\,\stackrel{\wedge}{\vec k}=
-\frac{1}{2}(\varphi\vec\tau^T\partial+\bbar\varphi\vec\tau
\bar\partial)$, the expression for $\stackrel{\wedge}{M^2}$ may
be written in the form
\begin{displaymath}
\stackrel{\wedge}{M^2}=-4\left[\frac{1}{2}(\stackrel{\wedge}
{\vec k^2}-\stackrel{\wedge}{\vec\imath^{\,2}})+\frac{1}{8}
(\stackrel{\wedge}{N^2}-\stackrel{\wedge}{Y^2})+2\stackrel
{\wedge}{N}+8\right].
\end{displaymath}
Essentially, in this case the operator $\stackrel{\wedge}{M^2}$
can only be expressed in terms of generators of the isotopic group
$SL_i(2,{\bf C})\otimes U_i(1)\otimes H_i(1)$; in so doing,
generators of the Lorentz group $SL_\ell (2,{\bf C})\otimes U_\ell
(1)\otimes H_\ell(1)$ do not occur. As a consequence of this
circumstance, the operator $\stackrel{\wedge}{M^2}$ has the same
expression on both the whole space ${\bf F}$ and subspace ${\cal
F}_0$.  Therefore, in the future we can consider either
$\stackrel{\wedge}{M^2}$ on the space ${\cal F}_0$ or (that is
equivalent) to pass into a Lorentzian system such that
$\varphi_{1k}=0$.

      The expression for $\stackrel{\wedge}{M^2}$ can be developed
into
\begin{equation}
\stackrel{\wedge}{M^2}=4\stackrel{\wedge}{\vec\imath^{\,2}}-
\stackrel{\wedge}{N^2}-10\stackrel{\wedge}{N}-32-2\left(
\bigtriangleup -\frac{1}{4}\stackrel{\wedge}{Y^2}-\frac{1}{4}
\stackrel{\wedge}{N^2}-\stackrel{\wedge}{N}\right)
\label{pr12}
\end{equation}
where $\bigtriangleup=\stackrel{\wedge}{\vec\imath^{\,2}}+\stackrel
{\wedge}{\vec k^2}$ is a Casimir operator of the algebra
$sl_i(2,{\bf C})$. On ${\cal F}_0$ we  have
($\varphi_k=\varphi_{2k}$ are additional variables,
$\partial_k=\frac{\partial}{\partial \varphi_k}$)
\begin{displaymath}
\bigtriangleup-\frac{1}{4}\stackrel{\wedge}{Y^2}-\frac{1}{4}
\stackrel{\wedge}{N^2}=\frac{1}{2}\left[(\varphi\vec\tau^T
\partial)^2+(\bbar\varphi\vec\tau\bar\partial)^2\right]-
\frac{1}{2}\left[(\varphi\partial)^2+(\bbar\varphi\bar
\partial)^2\right]=
\end{displaymath}
\begin{displaymath}
=\frac{1}{2}\sum^4_{a=1}\left[\varphi
\stackrel{+}{\tau}^T_a\partial\ \varphi\stackrel{+}{\tau}^T_a
\partial+\bbar\varphi\stackrel{+}{\tau}_a\bar\partial\
\bbar\varphi\stackrel{+}{\tau}_a\bar\partial\right].
\end{displaymath}
Using the comleteness condition of matrices
$\stackrel{+}{\tau}_a$:
\begin{displaymath}
\sum^4_{a=1}\left(\stackrel{+}{\tau}_a\right)_{km}
\left(\stackrel{+}{\tau}_a\right)_{np}=2\tau_{kn}\tau_{pm}
\end{displaymath}
where $\tau=\left(\matrix{0&1\cr -1&0\cr}\right)$, we obtain
\begin{displaymath}
\bigtriangleup -\frac{1}{4}\stackrel{\wedge}{Y^2}-\frac{1}{4}
\stackrel{\wedge}{N^2}=(\varphi\tau)_k\partial_m\varphi_k
(\partial\tau)_m+(\bbar\varphi\tau)_k\bar\partial_m\bbar
\varphi_k(\bar\partial\tau)_m=
\end{displaymath}
\begin{displaymath}
=\varphi\tau\varphi\ \partial\tau\partial +\bbar
\varphi\tau\bbar\varphi\ \bar\partial\tau
\bar\partial +\varphi\partial +\bbar\varphi\bar\partial\,.
\end{displaymath}
And as $\varphi\tau\varphi=0$, we have
\begin{displaymath}
\bigtriangleup -\frac{1}{4}\stackrel{\wedge}{Y^2}-\frac{1}{4}
\stackrel{\wedge}{N^2}=\stackrel{\wedge}{N}\,.
\end{displaymath}
Hence, the expression in round brackets in (\ref{pr12}) is equal
to zero. So that actually
\begin {displaymath}
\stackrel{\wedge}{M^2}=4\stackrel{\wedge}{\vec\imath^{\,2}}-
\stackrel{\wedge}{N^2}-10\stackrel{\wedge}{N}-32\,.
\end{displaymath}
Now it is not difficult to find the expression for $F_\Sigma(-
{\cal P}^2)$ in the equation $\stackrel{\wedge}{M^2}f^\Sigma=F_\Sigma
(-{\cal P}^2)F^\Sigma$ where $f^\Sigma=O^\Sigma w$ ($O^\Sigma$ is
the skeleton, $w$ is the Gibbs  distribution function). In the
coordinate representation we have
\begin{equation}
F_\Sigma(\Box)=F^0_\Sigma+\frac{N+6}{\mu^2}\Box
-\frac{\Box^2}{4\mu^4}
\label{pr13}
\end{equation}
where $F^0_\Sigma =-(N+5)^2-7+4i(i+1)$,  and $N$ and $i$ are the
isotonic number and isospin of the skeleton $O^\Sigma$
respectively.

\section {}
\label{dix2}
      There we find explicit amplitudes $O^\Sigma({\cal P,Q})$
determined by formula (\ref{eq611}):
\begin{eqnarray}
O^\Sigma({\cal P,Q}) =\int d^4\tilde\nu\,O^\Sigma(\varphi_{\alpha
k}\,,\varphi_k)\,e^{-\frac{\pi\rho}{2\mu^2}-2i\,\Im\!
\mbox{\scriptsize\it m}\,\bbar z_k\varphi_k}=\nonumber\\
=\exp\left(-\frac{-{\cal P}^2}{4\mu^2}\right)\,O^\Sigma
\left(-\frac{\partial}{\partial\bbar z},\frac{\partial}
{\partial z}\right)\,I
\label{pr21}
\end{eqnarray}
where $O^\Sigma (\varphi)$ are skeletons of particles, and
\begin{equation}
I(\pi\,,\rho)=\int d^4\tilde\nu\,\exp (\bbar\varphi_{\alpha k}
z_{\alpha k}-\bbar z_{\alpha k}\varphi_{\alpha k})
\label{pr22}
\end{equation}
such that $z_{\alpha k}=\delta_{\alpha 2} z_k$, and the measure
$d^4\tilde\nu$ is defined by formula (\ref{eq66}).

      Consider at first the integral appearing in the model
$h^{(*)}_8$ (see (\ref{eq36}))
\begin{displaymath}
{\cal J}=\int_0^{2\pi}\frac{d\omega}{2\pi}\,e^{\bbar\varphi_\alpha
z_\alpha-\bbar z_\alpha\varphi_\alpha}\,,\quad z_\alpha=z
\delta_{\alpha 2}
\end{displaymath}
where $\omega=\arg\varphi_2$. From definitions $\pi_\mu=\bbar\varphi
\stackrel{+}{\sigma}_\mu\varphi\,,\,\rho_\mu=\bbar z\stackrel{-}
{\sigma}_\mu z\ (\pi^2_\mu=\rho^2\mu=0)$ it follows that $\pi\rho=
2\vert\bbar z_\alpha\varphi_\alpha\vert^2$. As $2\pi\rho=-
{\cal P}^2$, $\bbar z_\alpha\varphi_\alpha=\frac{1}{2}\sqrt
{-{\cal P}^2}e^{i\chi}$. Therefore
\begin{displaymath}
{\cal J}=\frac{1}{2\pi}\int_0^{2\pi}e^{i\sqrt{-{\cal P}^2}
\sin\chi}\,d\omega\,.
\end{displaymath}
Since
\begin{displaymath}
\varphi_\alpha=e^{i\omega}{e^{i\psi}\vert\varphi_1\vert\choose
\vert\varphi_2\vert},\quad z_\alpha=e^{i\nu}{e^{i\alpha}\vert
z_1\vert\choose\vert z_2\vert},
\end{displaymath}
we obtain
\begin{displaymath}
\bbar\varphi_\alpha z_\alpha-\bbar z_\alpha\varphi_\alpha=2\left[
\sin(\omega-\nu +\psi-\alpha)\,\vert\varphi_1 z_1\vert+\sin(\omega-
\nu)\,\vert\varphi_2 z_2\vert\right],
\end{displaymath}
\begin{displaymath}
\bbar\varphi_\alpha z_\alpha +\bbar z_\alpha\varphi_\alpha=2\left[
\cos(\omega-\nu+\psi-\alpha)\,\vert\varphi_1 z_1\vert+\cos(\omega-
\nu)\,\vert\varphi_2 z_2\vert\right].
\end{displaymath}
It follows from this that
\begin{displaymath}
d\omega\,\frac{d}{d\omega}\sqrt{-{\cal P}^2}\,\sin\chi=\sqrt{-{\cal
P}^2}\,\cos\chi\,\frac{d\chi}{d\omega}\,d\omega=\sqrt{-{\cal P}^2}\,
\cos\chi\,d\omega\,,
\end{displaymath}
i.\ e.\ $d\chi=d\omega$, and consequently
\begin{displaymath}
{\cal J}=\frac{1}{2\pi}\int_0^{2\pi}e^{i\sqrt{-{\cal P}^2}
\sin\chi}\,d\chi={\cal J}_0(-{\cal P}^2)
\end{displaymath}
where ${\cal J}_0$ is the Bessel function.

      Integrating now (\ref{pr21}) with respect to $\omega_1$ and
$\omega_2$, we obtain
\begin{displaymath}
I=\int d^4\tilde\nu\,e^{\bbar\varphi_{\alpha k}z_{\alpha k}
-\bbar z_{\alpha k}\varphi_{\alpha k}}=\int d^2\nu\,
{\cal J}_0 (\sqrt{-{\cal P}^2_1})\,{\cal J}_0 (\sqrt
{-{\cal P}^2_2})
\end{displaymath}
where $-{\cal P}^2_2=2\rho^{(2)}\Pi\,,\,-{\cal P}^2_1=2\rho^{(1)}
(\pi-\Pi)$, and
\begin{displaymath}
d^2\nu=\frac{2}{\pi}\frac{\theta (\Pi_0)\theta (\pi_0-\Pi_0)}
{\theta(\pi_0)}\delta (\Pi^2)\,\delta (2\pi\Pi-\pi^2)\,d^4\Pi\,.
\end{displaymath}
Expanding functions ${\cal J}_0$ in a series, we shall have
\begin{displaymath}
I=\sum^\infty_{n,m=0}\frac{(-1)^n}{2^n(n!)^2}\ \frac{(-1)^m}
{2^m(m!)^2}\int d^2\nu\,\left(\rho^{(2)}\Pi\right)^n\,\left(
\rho^{(1)}(\pi-\Pi)\right)^m\,.
\end{displaymath}
As
\begin{displaymath}
\left(\rho^{(1)}(\pi-\Pi)\right)^m=\sum^m_{l=0}\left(\rho^{(1)}\pi
\right)^{m-l}\,\left(\rho^{(1)}\Pi\right)^l\,(-1)^l\,C^m_l\,,
\end{displaymath}
($C_l^m$ is the number of combinations),
\begin{displaymath}
I=\sum_{n,m,l}\frac{(-1)^{n+m+l}}{2^n(n!)^2 2^m(m!)^2}\,C^m_l
\left (\rho^{(1)}\pi\right)^{m-l}\,\int d^2\nu\,\left(\rho^{(1)}
\Pi\right)^l\,\left(\rho^{(2)}\Pi\right)^n\,.
\end{displaymath}
We need to calculate integrales in the form ($\pi^2=0$):
\begin{displaymath}
\int d^2\nu\,\overbrace{\Pi_\alpha\ldots\Pi_\omega}^s=\frac{1}
{\pi}\,\int\overbrace{\Pi_\alpha\ldots\Pi_\omega}^s\,\frac{\theta
(\Pi_0)\theta(\pi_0-\Pi_0)}{\theta_0}\,\delta (\Pi^2)\,\delta(\pi
\Pi)\,d^4\Pi\,.
\end{displaymath}
It is clear that the integration result must be of the form
$A(s)\overbrace{\pi_\alpha\ldots\pi_\omega}^s$ such that it
remains only to find $A(s)$. For this purpose we pass to the
system in which $\pi_\mu=(0,0,\pi_3,\pi_0)$ such that
$\pi_3=\pi_0$. In it we have
\begin{displaymath}
\frac{1}{\pi}\,\int\frac{\theta (\Pi_0)\theta (\pi_0-\Pi_0)}
{\theta_0}\,\delta (\Pi^2_1+\Pi^2_2+\Pi^2_3-\Pi^2_0)\,\delta
\left(\pi_0(\Pi_3-\Pi_0)\right)\,\times
\end{displaymath}
\begin{displaymath}
\times\,\Pi_\alpha\ldots\Pi_\omega\, d\Pi_1\, d\Pi_2\,
d\Pi_3\,d\Pi_0=\frac{1}{\pi}\,\int\delta (\Pi_1^2+
\Pi_2^2)\,d\Pi_1\, d\Pi_2\,\int^{\pi_0}_0\frac{d\Pi_0}{\pi_0}\,
\Pi_\alpha\ldots\Pi_\omega\,.
\end{displaymath}
As
\begin{displaymath}
\frac{1}{\pi}\,\int\delta (\Pi_1^2+\Pi_2^2)\,d\Pi_1\,d\Pi_2=\frac{1}
{\pi}\,\int^{2\pi}_0 d\theta\,\int^\infty_0\delta (r^2)\,r\,dr=1,
\end{displaymath}
and the integral $\frac{1}{\pi_0}\int_0^{\pi_0}d\Pi_0\,\Pi_\alpha
\ldots\Pi_\omega$ is only nonzero for $\alpha,\ldots,\omega$,
equal to 3 or 0, and $\Pi_3=\Pi_0$,
\begin{displaymath}
\frac{1}{\pi_0}\,\int_0^{\pi_0}d\Pi_0\,\overbrace{\Pi_\alpha\ldots
\Pi_\omega}^s=\frac{1}{\pi_0}\,\int_0^{\pi_0}d\Pi_0\,\Pi^s_0=
\frac{\pi^s_0}{s+1}.
\end{displaymath}
In any system the result is written as $\frac{1}{s+1}\overbrace
{\pi_\alpha\ldots\pi_\omega}^s$. Therefore for $A(s)$ we have
\begin{displaymath}
A (s)=\frac{1}{s+1}.
\end{displaymath}
It may be written the equality
\begin{displaymath}
\int d^2\nu\left(\rho^{(1)}\Pi\right)^l\,\left (\rho^{(2)}\Pi
\right)^n=\frac{1}{l+n+1}\left(\rho^{(1)}\pi\right)^l\,\left(
\rho^{(2)}\pi\right)^n\,,
\end{displaymath}
so that for $I$ we obtain
\begin{displaymath}
I=\sum_{n,m,l}\frac{(-1)^{n+m+l}}{2^n (n!)^2 2^m (m!)^2}\,C^m_l
\frac{(\pi\rho^{(2)})^n(\pi\rho^{(1)})^m}{l+n+1}\,.
\end{displaymath}
As
\begin{displaymath}
\sum^m_{l=0}\frac{(-1)^l}{l+n+1}\,C^m_l=\frac{m!n!}{(m+n+1)!},
\end{displaymath}
\begin{displaymath}
I=\sum^\infty_{n,m=0}\frac{(-\pi\rho^{(1)}/2)^m}{m!}\ \frac
{(-\pi\rho^{(2)}/2)^n}{n!(m+n+1)!}=
\end{displaymath}
\begin{displaymath}
=\sum^\infty_{n=0}\frac{(-\pi\rho^{(1)}/2)^m}{m!}\,{\cal J}_{m+1}
(\sqrt{2\pi\rho^{(2)}})\,\left(\frac{2}{\sqrt{2\pi\rho^{(2)}}}
\right)^{m+1}
\end{displaymath}
where ${\cal J}_{m+1}$ is the Bessel function. To take advantage
of the formula \cite{15}
\begin{displaymath}
\sum^\infty_{m=0}\frac{(-1)^m}{m!}\,\left(\frac{t(z+t/2)}{z}
\right)^m\,{\cal J}_{m+1}(z)=\frac{z}{z+t}\,{\cal J}_1(z+t)
\end{displaymath}
in which it is necessary to put $z=\sqrt{2\pi\rho^{(2)}}\,,\,
\pi\rho^{(1)}=t(\sqrt{2\pi\rho^{(2)}}+t/2)$, i.\ e.\ $t=\sqrt{2\pi
\rho}-\sqrt{2\pi\rho^{(2)}}$ where $\rho=\rho^{(1)}+\rho^{(2)}$,
we find for $I$ the final expression
\begin{displaymath}
I(\pi,\rho)=\frac{2}{\sqrt{2\pi\rho}}\,{\cal J}_1(\sqrt{2\pi\rho})=
\frac{2}{\sqrt{-{\cal P}^2}}\,{\cal J}_1(\sqrt{-{\cal P}^2}),\quad
-{\cal P}^2=2\pi\rho\,.
\end{displaymath}
At $\rho_\mu=0$ (i.\ e.\ $z=0$) it follows from this that
\begin{displaymath}
\int d^4\nu=1.
\end{displaymath}
It is not difficult to show that definition (\ref {pr21}) results
in the factorized expression for amplitudes $O^\Sigma ({\cal P,Q})$
or $O^\Sigma (\pi,\rho)$:
\begin{equation}
O^\Sigma (\pi,\rho)=O_\Sigma^{(l)}(\pi,\rho)\,O_\Sigma^{(i)}(z)\,
N_\Sigma (X),\quad X=\sqrt{2\pi\rho}
\label{pr23}
\end{equation}
where $O_\Sigma^{(l)}$ and  $O_\Sigma^{(i)}$ are denoted
Lorentzian and isotopic factors. Let us find at first all three
factors in the model $h^{(*)}_8$. In this model $L$-skeletons
$O^\Sigma (\varphi)$ are written in the form
\begin{displaymath}
O^\Sigma (\varphi)=\overbrace{\varphi_\alpha\ldots\varphi_\chi}^a\,
\overbrace{\bbar\varphi_{\dot\alpha}\ldots\bbar\varphi_{\dot
\omega}}^b\,.
\end{displaymath}
Spinor  indices equal to 2 answer additional variables. For
definiteness (due to the complete fermion-antifermion asymmetry)
we hold that $a\geq b$. Denoting $a+b=N$ and $a-b=Y$, we rewrite
$O^\Sigma (\varphi)$ in the form
\begin{displaymath}
O^\Sigma (\varphi)=\overbrace{\varphi_\delta\ldots\varphi_\chi}^Y\,
\overbrace{\varphi_\alpha\bbar\varphi_{\dot\alpha}\ldots
\varphi_\omega\bbar\varphi_{\dot\omega}}^{N-Y}\,.
\end{displaymath}
As $\pi_\mu=\bbar\varphi\stackrel{+}{\sigma}_\mu\varphi$,
$\varphi_\alpha\bbar\varphi_{\dot\alpha}=\frac{1}{2}\bbar\pi_{\alpha
\dot\alpha}$ where $\bbar\pi=\pi_\mu\stackrel{-}{\sigma}_\mu$.
Therefore, by definition of  amplitudes $O^\Sigma (\pi,\rho)$ it
follows that
\begin{displaymath}
O^\Sigma (\pi,\rho)=e^{-\frac{\chi^2}{4\mu^2}}\,
O^\Sigma\left(-\frac{\partial}{\partial\bbar z^\alpha}\,,
\,\frac{\partial}{\partial
z^{\dot\alpha}}\right)\,{\cal J}_0(X)=
\end{displaymath}
\begin{displaymath}
=e^{-\frac{\chi^2}{4\mu^2}}\,\overbrace{\frac{1}{2}
\bbar\pi_{\alpha\dot\alpha}\ldots\frac{1}{2}
\bbar\pi_{\omega\dot\omega}}^{(N-Y)/2}\left(-\frac{\partial}
{\partial\bbar z^\delta}\right)\ldots\left(-\frac{\partial}
{\partial\bbar z^\chi}\right)\,{\cal J}_0(X).
\end{displaymath}
To write $\frac{\partial}{\partial\bbar z^\alpha}=(\bbar\pi z)_\alpha
\frac{1}{X}\frac{\partial}{\partial X}$ and to use the formula
\cite{15}
\begin{displaymath}
\left(\frac{1}{X}\frac{\partial}{\partial X}\right)^m\,\frac{1}{X^\nu}
\,{\cal J}_\nu (X)=(-1)^m\,\frac{1}{X^{\nu +m}}\,{\cal J}_{\nu+m}(X),
\end{displaymath}
we obtain (here $(\bbar\pi z)_\delta=\bbar\pi_{\delta 2}z$)
\begin{equation}
O^\Sigma (\pi,\rho)=e^{-\frac{\chi^2}{4\mu^2}}\,
\overbrace{(\bbar\pi z)_\delta\ldots (\bbar\pi z)_\chi}^Y\,
\overbrace{\frac{1}{2}\bbar\pi_{\alpha\dot\alpha}\ldots
\frac{1}{2}\bbar\pi_{\omega\dot\omega}}^{(N-Y)/2}\,
\frac{1}{X^Y}{\cal J}_Y(X).
\label{pr24}
\end{equation}
Here $Y\geq 0$. In the case of $b\geq a\ (Y\leq 0)$ this formula
would be written in the form
\begin{displaymath}
O^\Sigma (\pi,\rho)=e^{-\frac{\chi^2}{4\mu^2}}\,
\overbrace{(\bbar z\bbar\pi)_{\dot\delta}\ldots(\bbar z
\bbar\pi)_{\dot\chi}}^{\vert Y\vert}\,\overbrace{\frac{1}
{2}\bbar\pi_{\alpha\dot\alpha}\ldots\frac{1}{2}\bbar
\pi_{\omega\dot\omega}}^{(N-\vert Y\vert)/2}\,
\frac{(-1)^Y}{X^{\vert Y\vert}}{\cal J}_{\vert Y\vert}(X).
\end{displaymath}

      If to distinguish additional
variables in $O^\Sigma (\varphi)$, it is necessary to write
\begin{displaymath}
O^\Sigma (\varphi)=\varphi^{a_s}\,\bbar\varphi^{b_s}\,\overbrace
{\varphi_\alpha\ldots\varphi_\omega}^{a'}\,\overbrace{\bbar
\varphi_{\dot\alpha}\ldots\bbar\varphi_{\dot\chi}}^{b'}
\end{displaymath}
where $a'=a-a_s\,,\,b'=b-b_s$. Here $\varphi\dot =\varphi_2\,,\,
\bbar\varphi\dot =\bbar\varphi_2$ are Lorentzian scalars. From
$O^\Sigma(\varphi)$ it is possible to construct the $SL(2,{\bf
C})$-multiplet, realizing a finite-dimensional representation
$(s,\dot s)=(a'/2,b'/2)$ of the group $SL_\ell (2,{\bf C})$. A
canonical basis of such representations is written in the form
\begin{equation}
O^\Sigma (\varphi)=\varphi^a_2\,\bbar\varphi^b_2\, O^{(s_0,
s_1)}_{j,j_3}(\zeta)
\label{pr25}
\end{equation}
where (see \cite{16})
\begin{displaymath}
O^{(s_0,s_1)}_{j,j_3}(\zeta)=(-1)^{j-s_0}\,\sqrt{\frac{(j+s_0)!
(j-s_0)!}{(j+j_3)!(j-j_3)!}}\,\zeta^{s_0+j_3}\left(1+\vert\zeta
\vert\right)^{s_1-s_0-1}\,\times
\end{displaymath}
\begin{displaymath}
\times\,P^{(s_0-j_3,s_0+j_3)}_{j-s_0}\left(
\frac{\vert\zeta\vert^2-1}{\vert\zeta\vert^2+1}\right).
\end{displaymath}
Here $s_0=s-\dot s=\frac{a'-b'}{2}=\frac{1}{2}F\,,\,s_1=s+\dot s+1=
\frac{1}{2}(a'+b')+1=\frac{1}{2}D+1\,,\,s_0\leq j\leq s_1-1\,,\,-j
\leq j_3\leq j\,,\,\zeta=\varphi_1/ \varphi_2$, and $P^{(\alpha,
\beta)}_n$ are Jacobi polynomials. The functions $O^{(s_0,s_1)}_{j,j_3}
(\zeta)$ are normalized by the condition
\begin{displaymath}
\frac{i}{2\pi}\,\int\overline{O^{(s_0,-s_1)}_{j',j_3'}(\zeta)}\,
O^{(s_0, s_1)}_{j,j_3}(\zeta)\,d\zeta\wedge d\bar\zeta=
\frac{(j-s_0)!}{2j+1}\,\delta_{jj'}\,\delta_{j_3j_3'}
\end{displaymath}
(pay attention that the representation $(s_0,-s_1)$ is conjugate
to $(s_0, s_1)$). Functions (\ref{pr23}) can developed further so
($Y_s=Y-F$)
\begin{displaymath}
O^\Sigma (\varphi)=(-1)^{j-s_0}\,\sqrt{\frac{(j+s_0)!(j-s_0)!}
{(j+j_3)!(j-j_3)!}}\,\varphi_1^{s_0+j_3}\,\varphi_2^{Y_3+s_0-
j_3}\,(\bbar\varphi_2\varphi_2)^{b_3}\,\times
\end{displaymath}
\begin{displaymath}
\times\,(\bbar\varphi\varphi)^{s_1-s_0-1}\,P^{(s_0-j_3,s_0+
j_3)}_{j-s_0}\left(\frac{\bbar\varphi\sigma_3\varphi}{\bbar
\varphi\varphi}\right).
\end{displaymath}
At the same time the amplitude $O^\Sigma (\pi,\rho)$ is written in
the form
\begin{displaymath}
O^\Sigma (\pi,\rho)=(-1)^{j-s_0}\,\sqrt{\frac{(j+s_0)!(j-s_0)!}
{(j+j_3)!(j-j_3)!}}\,\left(\frac{\pi_0-\pi_3}{2}\right)^{b_3}\,
\pi_0^{s_1-s_0-1}\,\times
\end{displaymath}
\begin{displaymath}
\times\,P^{(s_0-j_3,s_0+j_3)}_{j-s_0}\left(\frac{\pi_3}{\pi_0}
\right)\,(\pi z)^{Y_s+s_0-j_3}_2\,e^{-\frac{X^2}{4\mu^2}}\,
\frac{1}{X^Y}\,{\cal J}_Y(X)
\end{displaymath}
having obviously a factorized structure.

      Now we go over to the model $h^{(*)}_{16}$.  In the system
where $\varphi_{1k}=0$ (or in the space ${\cal F}_0$) the
skeletons of particles are expressed in terms of additional
variables $\varphi_k\dot = \varphi_{2k}\,,\,\bbar\varphi_k\dot
=\bbar\varphi_{2k}$ (isotopic variables now play a spin role)
\begin{displaymath}
O^\Sigma (\varphi)=\overbrace{\varphi_a\ldots\varphi_f}^A\,
\overbrace{\bbar\varphi_g\ldots\bbar\varphi_z}^B
\end{displaymath}
in a way analogous to the model $h^{(*)}_8$.
Representation (\ref{pr25}) is now rewritten in the form
\begin{displaymath}
O^\Sigma (\varphi)=\varphi^A_2\bbar\varphi^B_2\,O^{(i_0,i_1)}_{i,i_3}
(\zeta)\,,\quad\zeta=\frac{\varphi_1}{\varphi_2}
\end{displaymath}
where $i_0=\frac{1}{2}(A-B)=\frac{1}{2}Y\,,\, i_1=\frac{1}{2}(A+B)+1=
\frac{1}{2}N+1\,,\,\frac{1}{2}Y\leq i\leq\frac{1}{2}N\,,\,-i\leq
I_3\leq i$ are weights of a representation of the isotopic group
$SL_i(2,{\bf C})$. In the same way  as earlier on we have
\begin{displaymath}
O^\Sigma (\varphi)=(-1)^{i-i_0}\,\sqrt{\frac{(i+i_0)!(i-i_0)!}
{(i+i_3)!(i-i_3)!}}\,\varphi_1^{i_0+i_3}\,\varphi_2^{i_0-i_3}\,
(\bbar\varphi\varphi)^{i_1-i_0-1}\,\times
\end{displaymath}
\begin {displaymath}
\times\,P^{(i_0-i_3,i_0+i_3)}_{i-i_0}\left(\frac{\bbar
\varphi\tau_3\varphi}{\bbar\varphi\varphi}\right).
\end{displaymath}
Expanding Jacobi polynomials in a series
\begin{displaymath}
P^{(\alpha,\beta)}_n(x)=\frac{1}{2^n}\,\sum^n_{m=0}C^{n+
\alpha}_m\,C^{n+\beta}_{n-m}\,(x-1)^{n-m}\,(x+1)^m\,,
\end{displaymath}
we rewrite the previous formula:
\begin{displaymath}
O^\Sigma (\varphi)=\omega^{(i_1,i_0)}_{i,i_3}\,
\varphi_1^{i_0+i_3}\,\varphi_2^{i_0-i_3}\,(\bbar\varphi
\varphi)^{i_1-i_0-1}\,\times
\end{displaymath}
\begin{displaymath}
\times\,\sum^{i-i_0}_{m=0}(-1)^m\,C_m^{i-i_3}\,C^{i+i_3}_{i-
i_0-m}\,(\bbar\varphi_1\varphi_1)^m\,(\bbar\varphi_2
\varphi_2)^{i-i_0-m}
\end{displaymath}
where $\omega^{(i_1,i_0)}_{i,i_3}=\sqrt{\frac{(i+i_0)!(i-i_0)!}
{(i+i_3)!(i-i_3)!}}$. Since calculating $O^\Sigma (\pi,\rho)$
it is necessary to put $\varphi_k=-\frac{\partial}{\partial
\bbar z_k}$ such that $-\frac{\partial}{\partial\bbar z_k}=\frac
{\pi_0-\pi_3}{2}z_kD$ where $D=-\frac{2}{X}\frac{d}{dX}$, and
$X=\sqrt{2\pi\rho}=\sqrt{2(\pi_0-\pi_3)\bbar z_k z_k}$, it is
possible to write
\begin{displaymath}
\varphi_1^{i_0+i_3+m}\varphi_2^{i-i_3-m}\to\left(\frac{\pi_0-
\pi_3}{2}\right)^{i+i_0}\,z_1^{i_0+i_3+m}\,z_2^{i-i_3-m}\,D^{i+i_0}\,.
\end{displaymath}
Therefore, as $\bbar\varphi_k\varphi_k=\frac{1}{2}(\pi_0-\pi_3)$,
we have
\begin{displaymath}
O^\Sigma (\pi,\rho)=e^{-\frac{X^2}{4\mu^2}}\,\omega^{(i_1,i_0)}_{i,
i_3}\,(i+i_3)!\,(i-i_3)!\,\left(\frac{\pi_0-\pi_3}{2}\right)^{i_1+
i_0-1}\,\times
\end{displaymath}
\begin{displaymath}
\times\,\sum^{i-i_0}_{m=0}(-1)^m\,\frac{(\partial/ \partial z_1)^m}
{M!} ,\frac{(\partial/ \partial z_2)^{i-i_0-m}}{(i-i_0-m)!}\,
\frac{z_1^{i_0+i_3+m}}{(i_0+i_3+m)!}\,\frac{z_2^{i-i_3-m}}{(i-i_3-m)!}
\,D^{i+i_0}\,I(X).
\end{displaymath}
Denote $v_k=\frac{\pi_0-\pi_3}{2}\vert z_k\vert^2D$. The validity of
the following formula is verified by the method of induction
\begin{displaymath}
\frac{1}{m!}\,\left(\frac{\partial}{\partial z_m}\right)^m\,
z_k^{\alpha +m}\,f(X)=z^\alpha_k\,:\,L^\alpha_m (v_k)\,:\,f(X)\,.
\end{displaymath}
Here $f(X)$ is any function depending only on $z_k$ by means of $X$,
and $L^\alpha_m$ are Laguerre polynomials. A normal product sign
$:\,:$ suggests that the degree $(v_k)^n$ is understood to
be $\left(\frac{\pi_0-\pi_3}{2}\vert z_k\vert^2\right)^nD^n$.
Therefore $O^\Sigma (\pi,\rho)$ may be written as
\begin{displaymath}
O^\Sigma (\pi,\rho)=e^{-\frac{X^2}{4\mu^2}}\,\omega^{(i_1,
i_0)}_{i,i_3}\,(i+i_3)!\,(i-i_3)!\,\left(\frac{\pi_0-\pi_3}
{2}\right)^{i_1+i_0-1}\,\times
\end{displaymath}
\begin{displaymath}
\times\,z_1^{i_0+i_3}\,z_2^{i_0-i_3}\,\sum^{i-i_0}_{m=0}(-1)^m\,\frac
{L_m^{i_0+i_3}(v_1)\,L^{i_0-i_3}_{i-i_0-m}(v_2)}{(i_0+i_3+m)!
(I-i_3-m)!}\,D^{i+i_0}\,I(X).
\end{displaymath}

      We cannot sum up here the series over any $v_1$ and $v_2$.
But this is unnecessary. Since in the theory the parameters $z_1$
and $z_2$ are large (see paragraph \ref {kd6}), it suffices to know
the result at $v_1, v_2\to\infty$. As $L^\alpha_n(x)\smash
{\mathop{\longrightarrow}\limits_{x\to\infty}}(-x)^n/n!$, for
large $v_1$ and $v_2$ we have
\begin{displaymath}
\sum^{i-i_0}_{m=0}(-1)^m\,\frac{L_m^{i_0+i_3}(v_1)\,L^{i_0-i_3}_{i-
i_0-m}(v_2)}{(i_0+i_3+m)!(i-i_3-m)!}\ \to
\end{displaymath}
\begin{displaymath}
(-v_2)^{i-i_0}\,\sum^{i-i_0}_{m=0}\left(-\frac{v_1}{v_2}\right)^m
\,\left[m!(i_0+i_3+m)!(i-i_0-m)!(i-i_3-m)!\right]^{-1}=
\end{displaymath}
\begin{displaymath}
=\frac{(-v_2)^{i-i_0}}{(i-i_0)!(i-i_3)!(i_0+i_3)!}\,_2\!F_1
(-i+i_0,-i+i_3;i_0+i_3+1;-v_1/v_2)=
\end{displaymath}
\begin{displaymath}
=\frac{(v_1+v_2)^{i-i_0}}{(i-i_3)!(i+i_3)!}\,P^{(i_0-i_3,
i_0+i_3)}_{i-i_0}\left(\frac{v_1-v_2}{v_1+v_2}\right).
\end{displaymath}
Here $_2\!F_1$ is the hypergeometric function, and $P^{(\alpha,
\beta)}_n$ are Jacobi polinomials. As $v_1+v_2=\frac{\pi_0-\pi_3}
{2}\bbar zzD$, and $\frac{v_1-v_2}{v_1+v_2}=\frac{\bbar z\tau_3 z}
{\bbar zz}$, finally for $O^\Sigma (\pi,\rho)$ we obtain
\begin{displaymath}
O^\Sigma (\pi,\rho)=e^{-\frac{X^2}{4\mu^2}}\,\omega^{(i_1,i_0)}_{i,
i_3}\,\left(\frac{\pi_0-\pi_3}{2}\right)^{i_1+i_0-1}\,z_1^{i_0+i_3}
\,z_2^{i_0-i_3}\,(\bbar zz)^{i-i_0}\,\times
\end{displaymath}
\begin{displaymath}
\times\,P^{(i_0-i_3,i_0+i_3)}_{i-i_0}\left(\frac{\bbar
z\tau_3 z}{\bbar zz}\right)\,D^{2i}\,I(X).
\end{displaymath}
Thus, we again have come to Jacobi polynomials from which we
proceeded. Next, as
\begin{displaymath}
D^{2i}\,I=\left(-\frac{2}{X}\frac{d}{dX}\right)^{2i}\,
\frac{2}{X}\,{\cal J}_1(X)=\left(\frac{2}{X}\right)^{2i+1}\,
{\cal J}_{2i+1}(X),
\end{displaymath}
see \cite{15}, if to keep in mind that on the space ${\cal F}_0$
the equality $\frac{\pi_0-\pi_3}{2}=\frac{X^2}{4\bbar zz}$ is taken
place, we shall have
\begin{displaymath}
O^\Sigma (\pi,\rho)=e^{-\frac{X^2}{4\mu^2}}\,\omega^{(i_1,i_0)}_{i,
i_3}\,\frac{z_1^{i_0+i_3}z_2^{i_0-i_3}}{(\bbar zz)^{i_1+i_0-1}}\,
\times
\end{displaymath}
\begin{displaymath}
\times\,P^{(i_0-i_3,i_0+i_3)}_{i-i_0}\left(\frac{\bbar z\tau_3z}
{\bbar zz}\right)\,\left(\frac{X}{2}\right)^{2i_1-3}\,{\cal J}_{2i+
1}(X).
\end{displaymath}
This expression takes place when a representation of the algebra
$h^{(*)}_{16}$ is considered on the subspace ${\cal F}_0$, i.\
e.\ when $\varphi_{1k}=0$. In the case the additional variables
$\varphi_{2k}$ and spinors $\varphi_{\alpha,k}$ have no
distinctions.  It is not difficult to pass to the case when a
representation of $h^{(*)}_{16}$ is considered on the whole space
${\bf F}={\cal F}_F\otimes {\cal F}_0$. It is only necessary
that not all $\frac{\pi_0-\pi_3} {2}$ are equal to $\frac{X^2}{4\bbar
zz}$, namely, $O^\Sigma (\pi,\rho)$ is written so:
\begin{eqnarray}
O^\Sigma (\pi,\rho)=\left(\frac{\pi_0-\pi_3}{2}\right)^F\,
\omega^{(i_1,i_0)}_{i,i_3}\,\frac{z_1^{i_0+i_3}z_2^{i_0-i_3}}
{(\bbar zz)^{i_1+i_0-F-1}}\,P^{(i_0-i_3,i_0+i_3)}_{i-i_0}
\left(\frac{\bbar z\tau_3 z}{\bbar zz}\right)\,\times\nonumber\\
\times\,\left(\frac{X}{2}\right)^{2i_1-2F-3}
\,{\cal J}_{2i+1}(X)\,e^{-\frac{X^2}{4\mu^2}}\qquad
\label{pr26}
\end{eqnarray}
where $F$ is the fermion charge of amplitude. The rest
$\frac{\pi_0- \pi_3}{2}$ is the second component of spinor
$\varphi_\alpha
(\pi)=\frac{1}{2}{\pi_1+i\pi_2\choose\pi_0-\pi_3}=\frac{1}{2}\bbar
\pi_{\alpha 2}$ (the first component disappears when passing to
${\cal F}_0$ and occurs in the return transition from ${\cal F}_0$ to
${\bf F}$). Therefore on ${\bf F}$ the Lorentzian factor $\left(\frac
{\pi_0-\pi_3}{2}\right)^F$ in (\ref{pr23}) is written in the form
\begin{equation}
O^\Sigma_l=\frac{1}{2}\,\overbrace{\bbar\pi_{\alpha 2}\ldots
\bbar\pi_{\omega 2}}^F=\frac{1}{2^F}\,\varphi_\alpha (\pi)\ldots
\varphi_\omega (\pi)\,.
\label{pr27}
\end{equation}
For isotopic factor we shall have
\begin{equation}
O^\Sigma_i=\omega^{(i_1,i_0)}_{i,i_3}\,\frac{z_1^{i_0+i_3}
z_2^{i_0-i_3}}{(\bbar zz)^{i_1+i_0-F-1}}\,P^{(i_0-i_3,
i_0+i_3)}_{i-i_0}\left(\frac{\bbar z\tau_3 z}{\bbar zz}\right)\,,
\label{pr28}
\end{equation}
and for $N_\Sigma$ we obtain
\begin{equation}
N_\Sigma=\left(\frac{M_\Sigma}{2}\right)^{2i_1-2F-3}\,
{\cal J}_{2i+1}(M_\Sigma)\,e^{-\frac{M^2_\Sigma}{4\mu^2}}
\label{pr29}
\end{equation}
(let us assume $X^2=M^2_\Sigma$). As $\frac{\bbar z\tau_3z}
{\bbar zz}=\frac{\vert\eta\vert^2-1}{\vert\eta\vert^2+1}=\frac
{1}{\Lambda^2}=\frac{1}{136}$, in (\ref{pr28}) it is possible to
take $P^{(i_0-i_3,i_0+i_3)}_{i-i_0}(0)$.

      Consider the $L$-spinor $\varphi_\alpha (\pi)$. It may
be written as
\begin{displaymath}
\varphi_\alpha (\pi)=\frac{1}{2}\bbar\pi_{\alpha 2}=\bbar\pi a=
\frac{\stackrel{-}{\cal P}+\stackrel{-}{\cal Q}}{2}a
\end{displaymath}
where $a=\frac{1}{2}{0\choose 1}$ is a constant spinor. Define
now a $R$-spinor by the formula
\begin{displaymath}
\chi^{\dot\alpha}({\cal P,Q})=\frac{\stackrel{+}{\rho}\bbar\pi a}
{X}=\frac{\left(\stackrel{+}{\cal Q}-\stackrel{+}{\cal P}
\right)\left(\stackrel{-}{\cal P}+\stackrel{-}{\cal Q}\right)a}
{4X}\,.
\end{displaymath}
If in (\ref{eq45}) instead of $O^\Sigma ({\cal P,Q})$ we
substitute $\psi={\chi^{\dot\alpha}({\cal
P,Q})\choose\varphi_\alpha({\cal P,Q})}$ and subsequantly we
integrate on ${\cal Q}$, previously having put $Y_\mu=0$ (thus
linear powers of ${\cal Q}$ vanish), we come to a bispinor
$\psi={\chi\choose\varphi}$ equal to
\begin{displaymath}
\psi({\cal P})=\frac{1}{2}{M_\Sigma a\choose\stackrel{-}
{\cal P}a}.
\end{displaymath}
As can be seen, a wave function of a Dirac particle in the
momentum representation is a Penrose twistor.

\section{}
      Here it will be shown that the number $\Lambda$ including
in the commutation relations (\ref{eq12}) cannot be any number, it
is connected with the dimension ${\rm dim}\,d$ of  the Lie algebra
$d$ of the dynamical group by the relation $\Lambda=\sqrt{{\rm
dim}\,d}$.

      Consider the general case of the algebra $h_{2n}$; the dimension
of its automorphism group $Sp(n, {\bf C})$ (the dynamical group
${\cal D}^{(n)}$ for which the Lie algebra is denoted by
$d^{(n)}$) is equal, as is known, to ${\rm dim}\,d^{(n)}=\frac{N
(N+1)}{2}$ where $N=2n$.

      First of all it should be noticed that if representations
with different $\Lambda$ and $\Lambda'$ are mathematically
equivalent, satisfying the condition $\Lambda\Lambda'> 0$ (they
are not equivalent to the representation with $\Lambda=0$), then
physically such representations are not equivalent since different
electromagnetic charges are connected with them (as will be shown
below).

      Consider an arbitrary transformation (an excitation of
system) from the reductive group of dynamical transformations
${\cal D}^n\otimes U(1)$ generated by the Lie algebra
$d^{(n)}\oplus\frac{\Lambda}{4}$ (its dimension is equal to
$\frac{N(N+1)}{2}+1$). Since the semisimple group ${\cal D}^{(n)}$
covers twice the group $Sp(n,{\bf C})$ to realize a ``spinor''
representation of the latter (just consider the subgroup
$SL(2,{\bf    C})$ in $Sp(n,{\bf C})$ which is covered twice by an
appropriate subgroup in ${\cal D}^{(n)}$, see \cite{4};
in publications the representation is referred to as a metaplectic
one), and quadruply the minimal group $Sp(n,{\bf C})\!/\!{\bf
Z}_2$ which is locally isomorphic to the group $Sp(n,{\bf C})$
(see \cite{1} where the group ${\cal D}^{(n)}$ realizes
semispinor representations of the rotation group $SO(3)$), it is
natural to consider that the reductive group covers quadruply the
minimal group locally isomorphic to it. Therefore the group $u(1)$
covers quadruply the group $U(1)$, a generator of which is
$\Lambda$, and consequently $\frac{\Lambda}{4}$ is a generator of
the group $u(1)$.

      A transformation from the group ${\cal D}^n\otimes u(1)$ is
written in the form $e^{i\varphi}$ where
$\varphi=\Gamma_k\theta_k+ \frac{\Lambda}{4}\theta$. In the case
one can write an eikonal equation:
\begin{equation}
\sum^{n
(2n+1)}_{k=1}\left(\frac{\partial\varphi}{\partial\theta_k}
\right)^2+\left(\frac{\partial\varphi}{\partial\theta}\right)^2=
\sum^{n(2n+1)}_{k=1}\Gamma^2_k+\left(\frac{\Lambda}{4}\right)=0\,.
\label{pr31}
\end{equation}
Hence we obtain
\begin{equation}
\vert\Lambda\vert=\left(-16\,\sum^{n(2n+1)}_{k=1}\Gamma^2_k
\right)^{1/2}\,.\label{pr32}
\end{equation}
Now we shall show that
$-16\,\sum^{n(2n+1)}_{k=1}\Gamma^2_k=n(2n+1)$.

      Let a set of $2n\times 2n$-matrices $\{\gamma_k\},\,k=1,2,
\ldots,n(2n+1)$ be a basis in $Sp(n)$. These matrices are
antisymmetric in the sense that they satisfy the condition
$E^{-1}\gamma_k^T E=-\gamma_k$. They obey to Lie brackets
$[\gamma_k,\gamma_m]=C^{\,l}_{km}\gamma_l$ where $C^{\,l}_{km}$
are structural constants of the Lie algebra $sp(n)$.  The mapping
$\gamma_k\to\Gamma_k=\frac{1}{2\Lambda}E\phi\gamma_k\phi$ where
$\phi_\alpha$ are generators of the algebra $h_{2n}$ satisfying
the commutation relations
\begin{equation}
[\phi_\alpha,\phi_\beta]=\Lambda E_{\alpha\beta}
\label{pr33}
\end{equation}
defines  a homomorphism (in fact an isomorphism) from the algebra
$sp(n)$ to the algebra $d^{(n)}$, as
\begin{equation}
[\Gamma_k,\Gamma_m]=\frac{1}{2\Lambda}E\phi[\gamma_k,\gamma_m]
\phi=C^{\,l}_{km}\Gamma_l\,.
\label{pr34}
\end{equation}

      Consider the complete orthonormal system of $2n\times
2n$-matrices $\gamma_k$ acting in $h_{2n}$ as in a linear
space. The normalization condition is written in the form
\begin{equation}
(\gamma_k,\gamma_m)=Sp\,\gamma_k\gamma_m=c\delta_{km}\,.
\label{pr35}
\end{equation}
These matrices can be chosen in such  a way that the antisymmetric
ones coincide with a basis of the Lie algebra $sp(n)$,
$c=\frac{1}{2}$ \cite{17}.

      From (\ref{pr35}) it follows that if $\gamma_k\neq 1$, $Sp\,
\gamma_k=0$. For the unit matrix we fix the index $k=0$. From
(\ref{pr35}) it follows that $\gamma_0=\sqrt{\frac{c}{2n}}I$ where
$I$ is the $2n\times 2n$-unity  matrix. We write now the
completeness condition of matrices $\gamma_k$:
\begin{equation}
\sum^{(2n)^2-1}_{k=0}(\gamma_k)_{\alpha\beta}\,
(\gamma_k)_{\gamma\delta}=c\,\delta_{\alpha\delta}\,
\delta_{\gamma\beta}\,.
\label {pr36}
\end{equation}
This condition is in agreement with the normalization condition
(\ref{pr35}).

      Let us contract formula (\ref{pr36}) with generators
$\phi$ of the algebra $h_ {2n} $. As a result we obtain
\begin{equation}
\sum^{n(2n+1)}_{k=0}\left(E\phi\gamma_k\phi\right)\,
\left(E\phi\gamma_k\phi\right)=c\,E\phi\,\left(\phi E\phi
\right)\,\phi=-c\Lambda^2n^2\,,
\label{pr37}
\end{equation}
as $\phi E\phi=\Lambda n$ (it follows from (\ref{pr33})). In this
formula from the full set of $(2n)^2$ matrices $\gamma_k$
the only  remainders are $n(2n+1)$ antisymmetric matrices, forming
a basis of the Lie algebra $sp(n)$, and the unit matrix. Other
$n(2n-1)-1$ symmetric matrices (except for the unit one)
satisfying the condition $E^{-1} \gamma_k^TE=\gamma_k$ after
contracting with $\phi$ have given zero.  In the case of matrices
$\gamma_0$ we have
\begin{displaymath}
E\phi\gamma_0\phi=-\Lambda\sqrt{\frac{cn}{2}}\,.
\end{displaymath}
Since the matrix $\gamma_0$ does not belong to the algebra $sp(n)$,
we exclude it from the sum (\ref{pr37}). As a result we obtain
\begin{equation}
\sum^{n(2n+1)}_{k=1}\left(E\phi\gamma_k\phi\right)\,
\left(E\phi\gamma_k\phi\right) =-c\Lambda^2n^2-c\Lambda^2
\frac{n}{2}=-\frac{c}{2}\Lambda^2n(2n+1)\,.
\label{pr38}
\end{equation}
To  recall the definition of operators $\Gamma_k$ and assume
$c=1/2$, we come to the relation (attention should be paid to a
minus sign in the right part)
\begin{equation}
\sum^{n(2n+1)}_{k=1}\Gamma^2_k=-\frac{1}{16}n(2n+1)\,.
\label{pr39}
\end{equation}
From (\ref{pr32}) we now obtain
\begin{equation}
\vert\Lambda\vert=\sqrt{n(2n+1)}=\sqrt{\frac{N(N+1)}{2}}\,.
\label{pr310}
\end{equation}
Henceforth we shall see that the primeval value of the Sommerfeld
fine structure constant $\alpha=\frac{1}{\Lambda^2}$. In the case
of the algebra $h^{(*)}_{16}\quad n=8$, therefore
$\Lambda=\sqrt{136}$.  This result could also be obtained from the
Heisenberg theory \cite{3} (for this purpose all is present
there). In our opinion Heisenberg was close to this idea.

      We  remind also about the Eddington work \cite{18} in which,
as a matter of fact, he has predicted formula (\ref{pr310}) in the
specific case $n=8$, and he has "improved'' it having put the
right part equal to $\sqrt{\frac{N(N+1)}{2}+1}$. However, he
expressed the doubling of the number $n$, i.\ e.\ $N=2n$ not
correctly.
\newpage
\section{}
\label{dix4}
\centerline{Table of  masses of ``bare'' fundamental hadrons (in
GeV).}
\smallskip
\vbox{\offinterlineskip \hrule
\halign{\vrule#&\strut\quad\hfil#\quad&\vrule#&
\strut\quad\hfil#\quad&\vrule#&\strut\quad\hfil#\quad&\vrule#&
\strut\quad\hfil#\quad&\vrule#&\strut\quad\hfil#\quad&\vrule#\cr
height2pt&\omit&&\omit&&\omit&&\omit&&\omit&\cr
&\hfil&&$N:\ \ i=\frac{1}{2}$&
&$\Lambda:\ \ i=0$&&$\Sigma:\ \ i=1$&
&$\Delta:\ \ i=\frac{3}{2}$&\cr
&$M_\Sigma$\hfil&&$Y=1$&
&$Y=0$&&$Y=0$&&$Y=1$&\cr
&\hfil&&$N=2n+1$&&$N=2n+2$&
&$N=2n+2$&&$N=2n+3$&\cr
height2pt&\omit&&\omit&&\omit&&\omit&&\omit&\cr
\noalign{\hrule}
height2pt&\omit&&\omit&&\omit&&\omit&&\omit&\cr
&$n$\hfil&&Theor.\,\vrule\,Exper.&&Theor.\,\vrule\,Exper.&
&Theor.\,\vrule\,Exper.&&Theor.\,\vrule\,Exper.&\cr
height2pt&\omit&&\omit&&\omit&&\omit&&\omit&\cr
\noalign{\hrule}
&0\ &&1,14\ \ \ 0,94\ &&1,15\ \ \ 1,12\ &
&1,22\ \ \ 1,19\ &&1,34\ \ \ 1,23\ &\cr
&1\ &&1,27\ \ \ \ ---\ \ &&1,31\ \ \ 1,33\ &
&1,36\ \ \ \ ---\ \ &&1,44\ \ \ \ ---\ \ &\cr
&2\ &&1,39\ \ \ 1,39\ &&1,43\ \ \ 1,40\ &
&1,47\ \ \ 1,38\ &&1,55\ \ \ 1,55\ &\cr
&3\ &&1,49\ \ \ 1,47\ &&1,53\ \ \ 1,52\ &
&1,57\ \ \ 1,58\ &&1,64\ \ \ 1,65\ &\cr
&4\ &&1,60\ \ \ 1,65\ &&1,63\ \ \ 1,60\ &
&1,67\ \ \ 1,67\ &&1,73\ \ \ 1,69\ &\cr
&5\ &&1,69\ \ \ 1,67\ &&1,72\ \ \ 1,69\ &
&1,76\ \ \ 1,76\ &&1,91\ \ \ 1,90\ &\cr
&6\ &&1,78\ \ \ 1,78\ &&1,81\ \ \ 1,82\ &
&1,84\ \ \ 1,84\ &&1,89\ \ \ 1,89\ &\cr
&7\ &&1,86\ \ \ 1,81\ &&1,89\ \ \ 1,87\ &
&1,92\ \ \ 1,92\ &&1,97\ \ \ 1,95\ &\cr
&8\ &&1,94\ \ \ 1,99\ &&1,97\ \ \ 2,01\ &
&1,99\ \ \ 2,00\ &&2,04\ \ \ \ ---\ \ &\cr
&9\ &&2,02\ \ \ 2,00\ &&2,05\ \ \ 2,08\ &
&2,06\ \ \ 2,03\ &&2,11\ \ \ \ ---\ \ &\cr
&10&&2,09\ \ \ 2,10\ &&2,12\ \ \ 2,11\ &
&2,13\ \ \ 2,10\ &&2,18\ \ \ 2,16\ &\cr
height2pt&\omit&&\omit&&\omit&&\omit&&\omit&\cr}
\hrule}
\medskip
\vbox{\offinterlineskip
\hrule
\halign{\vrule#&\strut\quad\hfil#\quad&\vrule#&
\strut\quad\hfil#\quad&\vrule#\,\vrule&\strut\quad\hfil#\quad&\vrule#&
\strut\quad\hfil#\quad&\vrule#&\strut\quad\hfil#\quad&\vrule#\cr
height2pt&\omit&&\omit&&\omit&&\omit&&\omit&\cr
&\hfil&&$\Xi:\ \ i=\frac{1}{2}$&
&$\varepsilon:\ \ i=0$&&$\rho:\ \ i=1$&
&$K^*:\ \ i=\frac{1}{2}$&\cr
&$M_\Sigma$\hfil&&$Y=-1$&
&$Y=0$&&$Y=0$&&$Y=1$&\cr
&\hfil&&$N=2n+5$&&$N=2n$&
&$N=2n+2$&&$N=2n+3$&\cr
height2pt&\omit&&\omit&&\omit&&\omit&&\omit&\cr
\noalign{\hrule}
height2pt&\omit&&\omit&&\omit&&\omit&&\omit&\cr
&$n$\hfil&&Theor.\,\vrule\,Exper.&&Theor.\,\vrule\,Exper.&
&Theor.\,\vrule\,Exper.&&Theor.\,\vrule\,Exper.&\cr
height2pt&\omit&&\omit&&\omit&&\omit&&\omit&\cr
\noalign{\hrule}
&0\ &&1,39\ \ \ 1,31\ &&0,73\ \ \ 0,70\ &
&0,72\ \ \ 0,75\ &&0,84\ \ \ 0,89\ &\cr
&1\ &&1,50\ \ \ \ ---\ \ &&0,82\ \ \ 0,78\ &
&0,85\ \ \ \ ---\ \ &&0,95\ \ \ \ ---\ \ &\cr
&2\ &&1,60\ \ \ 1,53\ &&0,93\ \ \ 0,98\ &
&0,96\ \ \ 0,98\ &&1,05\ \ \ \ ---\ \ &\cr
&3\ &&1,69\ \ \ 1,68\ &&1,03\ \ \ 1,02\ &
&1,07\ \ \ 1,10\ &&1,15\ \ \ \ ---\ \ &\cr
&4\ &&1,75\ \ \ 1,82\ &&1,12\ \ \ 1,08\ &
&1,16\ \ \ 1,17\ &&1,23\ \ \ \ ---\ \ &\cr
&5\ &&1,86\ \ \ \ ---\ \ &&1,21\ \ \ 1,27\ &
&1,25\ \ \ 1,25\ &&1,32\ \ \ 1,28\ &\cr
&6\ &&1,94\ \ \ 1,94\ &&1,29\ \ \ 1,28\ &
&1,34\ \ \ 1,31\ &&1,40\ \ \ 1,40\ &\cr
&7\ &&2,02\ \ \ 2,03\ &&1,37\ \ \ 1,30\ &
&1,42\ \ \ 1,41\ &&1,47\ \ \ 1,43\ &\cr
&8\ &&2,09\ \ \ 2,12\ &&1,45\ \ \ 1,42\ &
&1,49\ \ \ 1,54\ &&1,55\ \ \ 1,58\ &\cr
&9\ &&2,16\ \ \ \ ---\ \ &&1,52\ \ \ 1,51\ &
&1,57\ \ \ 1,60\ &&1,62\ \ \ 1,65\ &\cr
&10&&2,23\ \ \ 2,25\ &&1,60\ \ \ 1,67\ &
&1,66\ \ \ 1,66\ &&1,68\ \ \ 1,70\ &\cr
height2pt&\omit&&\omit&&\omit&&\omit&&\omit&\cr}
\hrule}
\begin{displaymath}
\mu^2=0,065\,,\quad k^{-1}=2\cdot 10^{-14}\,{\rm cm},\quad
khc=1\,{\rm GeV,}
\end{displaymath}
The experimental mass values are taken from \cite{19}.

      We have to note that masses of low-lying states (in
particular, of nucleons) are appreciably renormalized by switching
on interactions, the existence of which are connected with the
degeneration group of the state $\dot f(z)$. It is necessary to
keep in mind that quanta of degeneration fields (first of all,
$\pi,\eta,K$-mesons, photon, graviton) are not fundamental
particles, they are a new kind of elementary particles.

\end{document}